\documentclass[a4paper,11pt]{article}
\pdfoutput=1 

\usepackage{jheppub} 

\usepackage[T1]{fontenc} 
\newcommand{\ii}{\mathrm{i}}
\newcommand{\beq}{\begin{eqnarray}}
\newcommand{\eeq}{\end{eqnarray}}

\newcommand{\bmp}{\noindent\begin{minipage}{16cm}}
\newcommand{\emp}{\end{minipage}\vskip 7mm} 

\newcommand{\SU}{\mbox{SU}}
\newcommand{\SO}{\mbox{SO}}
\newcommand{\SP}{\mbox{Sp}}
\newcommand{\UU}{\mbox{U}}

\definecolor{pumpkin}{rgb}{1.0, 0.4, 0.0}

\title{Loop-Generated Neutrino Masses in Composite Higgs Models}

\author[a,b]{Giacomo Cacciapaglia}
\author[c]{and Martin Rosenlyst}

\affiliation[a]{Institut de Physique des 2 Infinis (IP2I),
CNRS/IN2P3, UMR5822, 69622 Villeurbanne, France}
\affiliation[b]{Universit\' e de Lyon, Universit\' e Claude Bernard Lyon 1, 69001 Lyon, France}
\affiliation[c]{CP$^3$-Origins, University of Southern Denmark, Campusvej 55, DK-5230 Odense M, Denmark\\}

\emailAdd{g.cacciapaglia@ipnl.in2p3.fr}
\emailAdd{rosenlyst@cp3.sdu.dk}

\abstract{We present a composite scotogenic model for neutrino masses, which are generated via loops of $\mathbb{Z}_2$--odd composite scalars. 
We consider three different approaches to the couplings of the neutrinos (including three right-handed singlets) and the composite sector: ETC-like four-fermion interactions, fundamental partial compositeness and fermion partial compositeness. In all cases, the model can feature sizeable couplings and remain viable with respect to various experimental constraints if the three $ \mathbb{Z}_2 $--odd right-handed neutrinos have masses between the TeV and the Planck scales. Additionally, the lightest $\mathbb{Z}_2$--odd composite scalar may play the role of Dark Matter, either via thermal freeze-out or as an asymmetric relic. This mechanism can be featured in a variety of models based on vacuum misalignment. For concreteness, we demonstrate it in a composite two-Higgs scheme based on the coset $\SU(6)/\SP(6)$.}

\begin{document} 
\maketitle
\flushbottom

\section{Introduction}

In the Standard Model (SM) of particle physics, the masses of fermionic matter fields are generated via renormalizable couplings to the Higgs field, 
named Yukawa couplings. They present a hierarchical structure, connected with the very different masses observed in the charged fermions.
Nevertheless, the only neutral fermions, neutrinos, have masses that are several orders of magnitude smaller than those of the charged fermions. In the early realizations of the Yukawa couplings~\cite{Weinberg:1967tq}, they were even thought to be massless, until the discovery of oscillations~\cite{Pontecorvo:1967fh} convinced the scientific community that they must carry a mass, although very small. The simplest solution to this puzzle is the see-saw mechanism~\cite{Minkowski:1977sc,Mohapatra:1979ia,Yanagida:1980xy}, based on the existence of very heavy new states that couple to neutrinos and the Higgs boson via large couplings. This requires a typical new scale $\Lambda_{\rm see-saw} \approx 10^{12}$~GeV.

A more mysterious missing piece in our understanding of the Universe is the presence of Dark Matter (DM): no particle in the SM can account for it, while it constitutes 85\% of the total mass today.  
One of the attempts to put together the smallness of neutrino masses and the presence of DM is the so-called radiative see-saw or scotogenic model~\cite{Ma:2006km}. Here, neutrino masses vanish at tree level, like in the original formulation of the SM.  They are generated at one-loop level via a coupling to a second Higgs doublet and right-handed neutrinos that are odd under an exact $ \mathbb{Z}_2 $ symmetry of the model. Because of the latter, the lightest one of the two will be stable and play the role of a particle DM candidate.

In this work, we explore the possibility of realizing the one-loop radiative seesaw mechanism in a composite multi-Higgs scheme, where all the scalars are generated by the condensation of a confining new strong force at the TeV scale. For previous work on loop-generated neutrinos masses in composite scenarios see Refs.~\cite{Appelquist:2002me,Appelquist:2003hn}. The main scope of compositeness is, in this case, to dynamically generate the electroweak symmetry breaking (EWSB), thus alleviating the naturalness problem in the Higgs sector. 
It is well established that composite Higgs scenarios have difficulty in generating large effective Yukawa couplings~\cite{Bellazzini:2014yua,Panico:2015jxa,Cacciapaglia:2020kgq}, for which a phase of near-conformal dynamics or walking~\cite{Holdom:1981rm} is needed. In turn, it is also difficult to generate the large hierarchy between neutrino masses and the top mass, which should be due to very different behaviours of the responsible operators during the walking phase. Previous work on compositeness for neutrinos can be found in Refs.~\cite{Carmona:2013cq,Carmona:2014iwa,Carmona:2015ena,Frigerio:2018uwx}.
Realizing the one-loop see-saw mechanism could, therefore, help composite models to generate a viable fermion mass spectrum.
As we will see later, the composite nature of the scalars appearing in the loop plays a crucial role in predicting a near-degenerate spectrum, which yields a further suppression of the loop-induced neutrino masses. Compared to a composite realisation of the standard see-saw mechanism, this feature allows to lower substantially the scale where the right handed neutrino masses are generated. This is important because, as already mentioned earlier, generating a large variety of scales is a challenge for composite Higgs models and their UV completions. As already mentioned, the main ingredient is the presence of two Higgs doublets (2HDs), one of which protected by a discrete symmetry. Both are assumed to arise as pseudo-Nambu-Goldstone bosons (pNGBs) from the spontaneous breaking of the global symmetry of the confining sector. As such, this mechanism can be applied to a variety of models based on strong dynamics: ``Composite (Goldstone) Higgs'' (CH)~\cite{Kaplan:1983fs}, ``partially Composite Higgs'' 
(pCH)~\cite{Galloway:2016fuo,Alanne:2017rrs,Alanne:2017ymh,Barducci:2018yer},
``Little Higgs''~\cite{ArkaniHamed:2001nc,ArkaniHamed:2002qx}, 
``holographic extra dimensions''~\cite{Contino:2003ve,Hosotani:2005nz}, 
``Twin Higgs''~\cite{Chacko:2005pe,Batra:2008jy,Barbieri:2015lqa,Low:2015nqa} and 
``elementary Goldstone Higgs models''~\cite{Alanne:2014kea,Gertov:2015xma}.
We stress again that the near degeneracy of the scalar masses is a natural feature in the composite version, and an advantage compared to the elementary version based on an inert second Higgs doublet.

We start with a model containing two composite Higgs doublets $ H $ and $ \eta $ that are, 
respectively, even and odd under a discrete $ \mathbb{Z}_2 $ symmetry. 
The Higgs doublet $ H $ is identified with the SM one, where 
its neutral component, $ H^0 $, develops the electroweak (EW) vacuum expectation value (VEV). 
The small masses of the left-handed neutrinos are generated by a one-loop radiative seesaw mechanism with the neutral component of $ \eta $ and three new right-handed neutrinos, $ N_{R,i} $, running in a loop as shown in Fig.~\ref{loopdiagram}. 
By demanding that the three right-handed neutrinos (SM gauge singlets) transform as 
$ N_{R,i} \rightarrow -N_{R,i} $ under the same  $ \mathbb{Z}_2 $, while all the other fermions are even, implies that
only $ H $ couples to the charged fermions, while $ \eta $ couples only to the neutrinos. 
The composite Higgs-Yukawa (HY) couplings are generated via effective 
operators involving the elementary SM fermions and the 
composite Higgs doublets. The size of each coupling will depend on the scaling dimension
of the operator generating them, and could be of order unity for the top and suppressed by
an Ultra-Violet (UV) scale for the other fermions.
The mass hierarchy $ m_t/ m_\nu \gtrsim  10^{12}   $ is  thus generated with alleviated tuning by a one-loop 
radiative seesaw mechanism. Compared to the elementary model of Ref.~\cite{Ma:2006km}, the composite scenario naturally generate a near-degenerate scalar spectrum, thus yielding an additional suppression in the loop result that buys a few orders of magnitude in the mass scale for the singlet neutrinos.

For concreteness, we consider a minimal composite 2HD model fulfilling the above requirements, the $\SU(6)/\SP(6)$ model of Refs.~\cite{Cai:2018tet,Cai:2019cow}. Our choice is bestowed on this coset because it can be easily generated in a simple gauge-fermion underlying theory~\cite{Cacciapaglia:2020kgq}.
Other composite 2HD models were discussed, for instance, in Refs.~\cite{Mrazek:2011iu,Bertuzzo:2012ya,Ma:2015gra,Ma:2017vzm,Cacciapaglia:2019ixa,Cai:2020njb}. For our template model, we will investigate three different approaches to generate the HY effective operators: (i) ``Extended Technicolor'' (ETC)-type four-fermion operators, (ii) ``Fundamental Partial Compositeness'', and (iii) ``Partial Compositeness''. These three approaches can give rise to sizeable HY coupling constants, while the right-handed neutrino masses are allowed to range from the scale of the lightest $\mathbb{Z}_2$--odd composite particle ($ \sim $TeV scale) to the Planck scale. 
Furthermore, the lightest of the $\mathbb{Z}_2$--odd composite scalars may provide a viable (a)symmetric dark matter candidate~\cite{Cai:2019cow}. 

Finally, we will investigate the experimental constraints for these three approaches from lepton flavour violating processes, the $ h\rightarrow \gamma \gamma $ decay, the EW precision tests, gauge boson decay widths, and the DM relic density. 

\section{Composite two-Higgs doublet models with $ \mathbb{Z}_2 $ symmetry}

\begin{figure}[t!]
	\centering
	\includegraphics[width=0.45\textwidth]{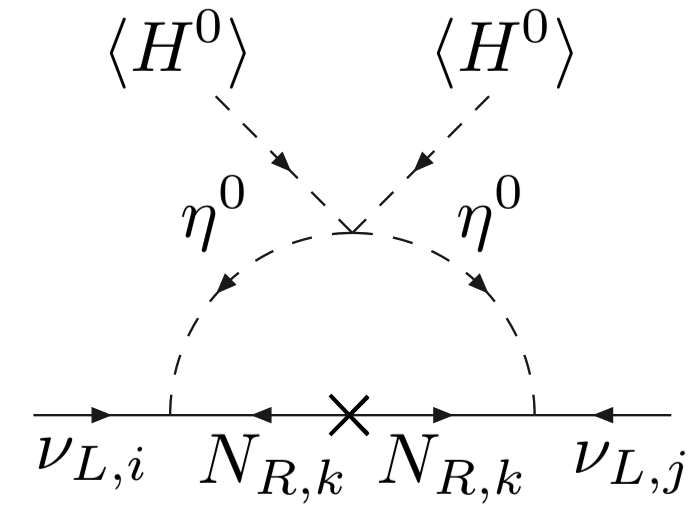}
	\caption{One-loop radiative Majorana neutrino mass in the scotogenic model proposed in Ref.~\cite{Ma:2006km}.}
	\label{loopdiagram}
\end{figure}

We presently focus on CH models with misalignment based on an underlying gauge description of strongly interacting fermions (hyper-fermions). The possible chiral symmetry breaking patterns in these CH models are discussed in Refs.~\cite{Witten:1983tx,Kosower:1984aw},
and we note the following minimal cosets with a Higgs candidate and custodial symmetry: SU(4)/Sp(4)~\cite{Galloway:2010bp}, SU(5)/SO(5)~\cite{Dugan:1984hq}, SU(6)/Sp(6)~\cite{Cai:2018tet}, SU(6)/SO(6)~\cite{Cacciapaglia:2019ixa}, and SU(4)$\times$SU(4)/SU(4)~\cite{Ma:2015gra}. Two composite Higgs doublets and a $\mathbb{Z}_2$ symmetry are present in the three latter cases~\cite{Cai:2018tet,Cacciapaglia:2019ixa,Ma:2017vzm}, where the coset $\SU(6)/\SP(6)$ generates the minimal number of pNGBs that simultaneously fulfils our requirements. This kind of model may also provide (a)symmetric dark matter candidates~\cite{Cai:2019cow}. Note that our proposal is rather general because the above requirements can also be fulfilled in other realisations that do not have a simple gauge-fermion underlying description, e.g. the models in Ref.~\cite{Mrazek:2011iu,Bertuzzo:2012ya}, and they can also be fulfilled in fundamental realisations as in Ref.~\cite{Ma:2006km}.

We remind the reader that Higgs naturalness is achieved in CH models as the EW scale is
generated dynamically via condensation of a new strong force, while vacuum misalignment~\cite{Contino:2010rs,Panico:2015jxa}
ensures that the pNGB Higgs is lighter than the compositeness scale and has SM-like couplings.
The underlying model consists of $ N_f $  Weyl hyper-fermions 
charged under a new strongly interacting ``hyper-color'' 
gauge group $ G_{\rm HC} $. The choice of either real, pseudo-real or a complex representation of the hyper-fermions under $ G_{\rm HC} $
determines the symmetry breaking pattern of the global symmetry $\mathrm{G} \to \mathrm{H}$. Finally,
the hyper-fermions are charged under the EW symmetry so that $\SU(2)_{\rm L} \times \UU(1)_{\rm Y}$ 
is contained in $\mathrm{H}$ when the vacuum is aligned to the EW preserving direction. 

This particular alignment, however, may not stable 
because of the presence of explicit breaking of $\mathrm{G}$ in 
the form of gauge interactions, fermion couplings (leading to the HY terms)  
and explicit masses for the hyper-fermions. This is the minimal set required for
having a realistic models, even though additional breaking terms may be generated
from the UV completion of the model. All in all, these terms generate a potential for
the pNGBs, which thus determines the ultimate alignment of the vacuum: while the
EW gauging and hyper-fermions masses typically tend to preserve the EW-preserving
direction, the top couplings are identified as the dynamical source for the EWSB.
The alignment of the vacuum, therefore, is moved away from the EW preserving one
by an angle $\sin \theta =  v_{\rm EW}/(2\sqrt{2}f)$~\cite{Kaplan:1983fs}, 
where $ v_{\rm EW} =246$~GeV and $f$ is the decay constant of the pNGBs 
depending on the confinement of the underlying strong dynamics. From 
EW precision measurements~\cite{Agashe:2006at,Grojean:2013qca}, 
this angle generically needs to be $\sin \theta \lesssim 0.2$, which also 
fixes $2\sqrt{2}f \gtrsim 1.2$~TeV. However, lower compositeness scales
may be allowed in specific cases~\cite{Ghosh:2015wiz,BuarqueFranzosi:2018eaj}.

In general, the large mass hierarchy between the top quark and neutrinos may be generated
in two ways: via very small neutrino couplings with respect to the top ones, or using highly
hierarchical VEVs for the 2HDs. In the following, we assume that the vacuum is only misaligned along the SM Higgs direction,
while the second Higgs doublet remains inert. This configuration can be stable once the fermion couplings are
properly chosen, i.e. the charged fermions couple to the first doublet only while neutrinos couple to 
the inert doublet and the right-handed neutrinos.
Majorana masses for the left-handed neutrinos are generated from a one-loop radiative seesaw mechanism with the one-loop Feynman diagram in Fig.~\ref{loopdiagram}. The coset structure can be schematically represented by a $N_f \times N_f$ matrix,
\beq
\left( \begin{array}{c|c}
\mathrm{G}_0/\mathrm{H}_0 & \begin{array}{c} \mathbb{Z}_2\mbox{--odd} \\ \mbox{pNGBs} \end{array} \\ \hline
\begin{array}{c} \mathbb{Z}_2\mbox{--odd} \\ \mbox{pNGBs} \end{array} & \begin{array}{c} \mathbb{Z}_2\mbox{--even} \\ \mbox{pNGBs} \end{array}
\end{array} \right), \label{eq. Nf Nf matrix}
\eeq
where $\mathrm{G}_0/\mathrm{H}_0$ is one of the two minimal cosets 
$\SU(4)/\SP(4)$ or $\SU(5)/\SO(5)$, with one composite Higgs doublet. The $ \mathbb{Z}_2 $ symmetry can be understood in terms of the underlying hyper-fermions 
$\psi_i$, $i = 1, \dots N_f$, that condense: $\psi_{5, \dots N_f}$ are $\mathbb{Z}_2\mbox{--odd}  $ while the hyper-fermions that participate to the minimal coset are $\mathbb{Z}_2\mbox{--even}  $. Among the $\mathbb{Z}_2\mbox{--odd}  $ pNGBs must be contained the $\mathbb{Z}_2\mbox{--odd}  $ Higgs doublet $ \eta $. 

\begin{table}[tb]
\begin{center}
{\renewcommand{\arraystretch}{1.2}
\begin{tabular}{c|c|c|c|c}
\hline\hline
    & $\mbox{G}_{\rm HC} $ &  $\mbox{SU(2)}_{\rm L} $ & $\mbox{U(1)}_{\rm Y} $ & $ \mathbb{Z}_2 $ \\
    \hline\hline
 $ \Psi_1\equiv (\psi_1,\psi_2)^T $  & $\Box$ &  $\Box$ &  $0$ &$ +1 $  \\ 
  $ \psi_3 $  & $\Box$ &  $\textbf{1}$&  $-1/2$ &$ +1 $  \\ 
  $ \psi_4 $  & $\Box$  &  $\textbf{1}$&  $+1/2$ &$ +1 $  \\
 $ \Psi_2\equiv (\psi_5,\psi_6)^T $  & $\Box$ &  $\Box$ &  $0$ &$ -1 $ \\
  $ N_{R,i} $  & $\textbf{1}$ & $\Box$ &  $0$ &$ -1 $  \\ 
\hline\hline
\end{tabular} }
\end{center}
\caption{The hyper-fermions and the right-handed neutrinos in the SU(6)/Sp(6) template model labelled with their representations of $ \rm G_{\rm HC} \times \SU(2)_L \times \UU(1)_Y $ and parity under the $ {\mathbb{Z}_2} $ symmetry. The index $ i=1,2,3 $ represents the generation number of the neutrinos. } \label{tab:fermionssu6sp6}
\end{table}

\section{A concrete composite 2HDM}

In the following, we focus on the  $\SU(6)/\SP(6)$ model~\cite{Cai:2018tet} as a template for this mechanism. We assume that four Weyl hyper-fermions are arranged in two $\SU(2)_{\rm L}$ doublets, $\Psi_1 \equiv (\psi_{1},\psi_{2})^T$ and $\Psi_2 \equiv (\psi_{5},\psi_{6})^T$, and in two $\SU(2)_{\rm L}$ singlets, $\psi_{3,4}$, with hypercharges $ \mp 1/2 $. In addition, we add three $\mathbb{Z}_2\mbox{--odd} $ right-handed neutrinos, $ N_{R,i} $, with the masses $ M_i $, which are SM gauge singlets. We have listed these fermions in Table~\ref{tab:fermionssu6sp6} with their representations of the gauge groups and their parity under the $ \mathbb{Z}_2 $ symmetry.

\subsection{The condensate and pNGBs}

The required symmetry breaking pattern can be achieved if the hyper-fermions are in a pseudo-real representation
of the confining group: this can be minimally achieved for $ \rm G_{\rm HC} = \rm \SU(2)_{HC} $ or $ \rm \SP(2N)_{\rm HC} $ with the hyper-fermions in the fundamental representation. The six Weyl fermions can be arranged into an $ \SU(6) $ vector $ \Psi \equiv (\psi^1,\psi^2,\psi^3,\psi^4,\psi^5,\psi^6)^T $. 
They form an anti-symmetric condensate in the form
\beq
\langle \Psi^I_{\alpha,a}\Psi^J_{\beta,b}\rangle\epsilon^{\alpha\beta}\epsilon^{ab}\sim \Phi^{IJ}_{\rm CH},
\eeq
where $ \alpha,\beta $ are spinor indices, $ a,b $ are HC indices, and $ I,J $ are flavour indices of the hyper-fermions. In the following, we will suppress the contractions of these indices for simplicity. 
A condensation in this operator visibly breaks $\SU(6) \to \SP(6)$ via an anti-symmetric tensor.

The CH vacuum of the model, giving rise to the EW VEV of $ H^0 $ by misalignment, can be written as~\cite{Galloway:2010bp} 
\beq \label{EW vacuum matrix}
\Phi_{\text{CH}}=\begin{pmatrix}i\sigma_2 c_\theta &\mathbf{1}_2 s_\theta&0\\ -\mathbf{1}_2 s_\theta &-i\sigma_2 c_\theta&0\\0&0&i\sigma_2\end{pmatrix},
\eeq 
where from now on we use the definitions $ s_x \equiv \sin x $, $ c_x\equiv \cos x $, and $ t_x \equiv \tan x $.

\begin{table}[tb]
\begin{center}
{\renewcommand{\arraystretch}{1.2}
\begin{tabular}{c|c|c}
\hline\hline
    & $\begin{array}{c} \mbox{EW vacuum} \\ (\theta = 0) \end{array}$ & $\begin{array}{c} \mbox{CH vacuum} \\ (\theta \neq 0) \end{array}$ \\
\hline\hline
$\displaystyle \mathrm{G}_0/\mathrm{H}_0$ &  $\begin{array}{c} H = (2,1/2)_+ \\ \chi = (1,0)_+ \end{array}$ & $\begin{array}{c} h, \; z^0, \; w^\pm \\ \chi \end{array}$ \\
\hline
$\begin{array}{c} \mathbb{Z}_2\mbox{--odd} \\ \mbox{pNGBs} \end{array}$ & $\begin{array}{c} \eta = (2,1/2)_- \\ \Delta = (3,0)_- \\  \varphi^0 = (1,0)_- \end{array}$ &   $\begin{array}{c} \mbox{Re}\ \eta^0,\; \mbox{Im}\ \eta^0 \; \eta^\pm \\ \Delta^0, \; \Delta^\pm \\  \varphi^0 \end{array}$ \\
\hline
$\begin{array}{c} \mathbb{Z}_2\mbox{--even} \\ \mbox{pNGBs} \end{array}$ & $\eta' = (1,0)_+ $ & $\eta'$ \\ 
\hline\hline
\end{tabular} }\end{center}
\caption{The pNGBs in the template SU(6)/Sp(6) model in the EW-preserving alignment, characterized by their $ (\rm SU(2)_L, U(1)_Y)_{\mathbb{Z}_2}$ quantum numbers, and in the CH vacuum. Note that $H = (w^+, (h+\ii z^0)/\sqrt{2})^T$, where $w^\pm, z^0$ are the Goldstones eaten by the $W^\pm$ and $Z$ bosons.} \label{tab:su6sp6}
\end{table}


The chiral symmetry breaking results in 14 pNGBs, $ \pi_a $ with $ a=1,...,14 $, corresponding to the broken generators, $ X_a $. 
Here we will work in the basis where the pNGBs are defined around the stable vacuum, so that none will be allowed to 
develop a VEV except for the composite SM Higgs candidate. Thus, we parameterize them as $\Sigma=\exp[i\pi_aX_a/f] \Phi_{\text{CH}}$, where $f$ is their decay constant. 
A preserved $ \mathbb{Z}_2 $ symmetry is identified in this vacuum, which can be written in terms of the following $ \SU(6) $ matrix: 
\beq 
P = \text{Diag}(1,1,1,1,-1,-1). \label{eq: parity symmetry}
\eeq 
A classification of the 14 pNGBs is provided in Table~\ref{tab:su6sp6} together with their parity assignment. We also provide the EW quantum numbers, which are only well defined in the EW-preserving vacuum, $\theta = 0$.
The $ \mathbb{Z}_2\mbox{--odd} $ pNGBs are, therefore, the second doublet $\eta$, a triplet $\Delta$ and a singlet $\varphi^0$. The right-handed neutrinos
will inherit an odd charge via their couplings to the strong sector, as we will illustrate below.

\subsection{The chiral Lagrangian and the effective potential}

In terms of the sixplet of Weyl spinors, $\Psi $, the underlying fermionic Lagrangian  can be written as 
\beq
\mathcal{L}_{\rm ferm.}= \Psi^\dagger i \gamma^\mu D_\mu \Psi - \frac{1}{2} \left(\Psi^T M_\Psi \Psi +{\rm h.c.} \right)
	 + \delta \mathcal{L}\,, 
\label{Basic Lagrangian (UV)}
\eeq
where the covariant derivatives include the $\rm G_{\rm HC}$ gluons and the 
$\SU(2)_{\rm L}$ and $ \UU(1)_{\rm Y} $ gauge bosons. The mass term consists of three independent masses
for the doublets and singlets hyper-fermions, $M_\Psi=\text{Diag}(im_1\sigma_2,-im_2\sigma_2,im_3\sigma_2)$. 
Note that for $m_1 = m_2 = m_3$, the mass matrix is proportional to the EW-preserving vacuum in Eq.~\eqref{EW vacuum matrix} with $\theta=0$: 
this is not by chance, as it is indeed the hyper-fermion masses that determine the signs in the vacuum structure~\cite{Cacciapaglia:2020kgq}.
The additional terms in $ \delta \mathcal{L}$ are interactions responsible for generating masses for the SM fermions
in the condensed phase, and we will illustrate their possible form in the following.

Below the condensation scale $ \Lambda_{\rm HC}\sim 4\pi f $, Eq.~\eqref{Basic Lagrangian (UV)} needs to be replaced by an effective Lagrangian:
\beq
    \label{eq:effLag}
    \mathcal{L}_\mathrm{eff}=\mathcal{L}_{\mathrm{kin}}-V_{\mathrm{eff}}, 
\eeq 
where $ \mathcal{L}_{\mathrm{kin}} $ is the usual leading order ($\mathcal{O} (p^2)$) chiral Lagrangian~\cite{Cacciapaglia:2020kgq}.  Besides providing kinetic terms and self-interactions for the pNGBs, it will induce masses for the EW gauge bosons and their couplings with the pNGBs (including the SM Higgs identified as $ h $), 
\beq \label{WZ masses and SM VEV}
&&m_W^2=2g_L^2f^2s_\theta^2,\quad\quad m_Z^2=m_W^2/c^2_{\theta_W}, \\
&&g_{hWW}=\sqrt{2}g^2_Wfs_\theta c_\theta=g_{  hWW}^{\rm SM}c_\theta,\quad\quad g_{ hZZ}=g_{ hWW}/c^2_{\theta_W}, \nonumber
\eeq 
where $ v_{\rm EW}\equiv 2\sqrt{2}fs_\theta = 246~\text{GeV} $, $ g_L $ is the weak $ \SU(2)_{\rm L} $ gauge coupling, and $ \theta_W $ is the Weinberg angle. 
The vacuum misalignment angle $ \theta $ parametrizes the 
corrections to the Higgs couplings to the EW gauge bosons and 
is constrained by LHC data~\cite{deBlas:2018tjm}. 
This would require a small $ \theta $ ($ s_\theta \lesssim 0.3 $), 
however an even smaller value is needed by the EW precision 
measurements ($s_\theta \lesssim 0.2$), as we will explain in Section~\ref{exp constraints}. 

The value of $\theta$, and the amount of misalignment, is controlled by 
the effective potential  $V_{\mathrm{eff}}$, which receives contributions from
 the EW gauge interactions, the SM fermion couplings 
to the strong sector, and the vector-like masses of the hyper-fermions. At leading order, each source of symmetry breaking contributes independently to
the effective potential in Eq.~(\ref{eq:effLag}):
\beq
V_{\text{eff}}&=&V_{\text{gauge}}+V_\text{m} +V_{\text{top}}+\dotsc, \label{Potential 1}
\eeq 
where the dots are left to indicate the presence of mixed terms at higher orders,
or the effect of additional UV operators.
In this work, we will write the effective potential in terms of effective operators,
which contain insertions of spurions that correspond to the symmetry breaking 
couplings. A complete classification of such operators, for this kind of cosets, up
to next-to-leading order can be found in~\cite{Alanne:2018wtp}.

Both the contribution of gauge interactions and of the hyper-fermion masses arise
to leading $\mathcal{O} (p^2)$ order and have a standard form:
\begin{equation}
\begin{aligned} V_{\rm gauge,p^2}&= C_g f^4 \sum_{i=1}^3 g_L^2 \text{Tr}[T_L^i \Sigma T_L^{iT}\Sigma^\dagger]+g_Y^{2}\text{Tr}[T_R^3 \Sigma T_R^{3T} \Sigma^\dagger] \\ &=- C_g f^4 \left(\frac{3g_L+g^{2}_Y}{2}c_\theta^2 - \frac{3g_L^2}{2}+\dots\right), \end{aligned} \end{equation} 
where $T_{L/R}$ are the gauged generators embedded in the global $\SU(6)$, while
\begin{equation}
\begin{aligned}  
V_{\rm m} = -2 \pi Z f^3 \text{Tr}[M_\Psi \Sigma^\dagger]+\text{h.c.}
\end{aligned}
\end{equation}
Here, $C_g$ and $Z$ are $ \mathcal{O}(1) $ form factors that can be computed on the lattice (e.g. $ Z\approx 1.5 $ in Ref.~\cite{Arthur:2016dir} for the $ \SU(2) $ gauge theory with two Dirac (four Weyl) hyper-fermions). Both these terms prefer the vacuum aligned with the EW-preserving direction, so that EWSB is crucially related to the SM fermion mass generation, or more specifically the top quark.\\

\textbf{Top quark mass:} 
Generating a large enough mass for the top quark is a well-known hurdle for all CH models~\cite{Bellazzini:2014yua,Panico:2015jxa,Cacciapaglia:2020kgq}.
The most traditional approach consists in adding bi-linear couplings of the top quark fields to the strong sector, in the form of four-fermion interactions generated by a gauge extension of the condensing gauge symmetry, ``Extended Technicolor'' (ETC)~\cite{Dimopoulos:1979es}. The main issue with this approach is the fact that the ETC scale needs to be low, thus also generating dangerous flavour changing neutral currents (FCNCs).
An alternative approach, revived in the holographic model, is that of ``Partial Compositeness'' (PC) proposed in Ref.~\cite{Kaplan:1991dc}, where the top quark features a linear coupling to the strong sector. This helps avoiding FCNCs and generating a large top mass via enhancement from large anomalous dimensions of the fermionic operators the top couples to.

In the underlying gauge-fermion model we consider here, the PC operators require the extension of the model in Table~\ref{tab:fermionssu6sp6} by a new specie of fermions $\chi_t$, transforming under the two-index anti-symmetric representation of $ G_{\rm HC}=\SP(2N)_{\rm HC} $, and carrying appropriate quantum numbers under the SM gauge symmetry. For the top, it is enough to introduce a vector-like pair with hypercharge $+2/3$ and transforming as a fundamental of QCD.  Models of this type were first proposed in Refs.~\cite{Barnard:2013zea,Ferretti:2013kya}, where our model is an extension of the one in Ref.~\cite{Barnard:2013zea}.

The four-fermion interactions generating PC operators can have various forms. In this work, we will be interested specifically on the following ones: 
\beq \label{eq: top PC operators}
\frac{\widetilde{y}_{L}}{\Lambda_t^2} Q^{\alpha\dagger}_{L} (\Psi^\dagger P_Q^\alpha \Psi^*\chi_{t}^\dagger) + \frac{\widetilde{y}_{R}}{\Lambda_t^2}t_{R}^{c\dagger} (\Psi^\dagger P_t \Psi \chi_{t}^\dagger)+\rm h.c., \ \ \ \ \ \
\eeq  
where $P_Q$ and $P_t$ are spurions that project-out specific combination of the flavour components in the sixplet $\Psi$.
By choice, for the left-handed top the spurions transform as the symmetric representation of the chiral symmetry $ \SU(6) $, while for the right-handed top the spurion transforms the adjoint representation. Moreover, we will impose that only the $\mathbb{Z}_2$--even hyper-fermions couple to the top fields, so that this parity remains preserved.\footnote{The $\mathbb{Z}_2$ may be broken by additional UV operators.} Concretely, the left-handed spurions are  given by the matrices
\begin{equation}
\begin{aligned}  
P_{Q,ij}^1=\frac{1}{\sqrt{2}}(\delta_{i1}\delta_{j3} + \delta_{i3}\delta_{j1}), \quad \quad P_{Q,ij}^2=\frac{1}{\sqrt{2}}(\delta_{i2}\delta_{j3} + \delta_{i3}\delta_{j2}), \end{aligned}  
\end{equation} 
while the right-handed spurion has three components, which can be written as 
\begin{equation}
\begin{aligned}   P_t=A_t P_t^1 +B_t P_t^2 + C_t P_t^3,   \end{aligned}  
\end{equation} 
where 
\begin{equation}
\begin{aligned}  & P_t^1=\frac{1}{\sqrt{2}}\text{Diag}(0,0,1,-1,0,0),\nonumber \\ & P_t^2=\frac{1}{2}\text{Diag}(1,1,-1,-1,0,0),\nonumber \\ \nonumber &P_t^3=\frac{1}{2\sqrt{3}}\text{Diag}(1,1,1,1,-2,-2). \end{aligned} 
\end{equation} 
These PC operators also generate the contributions, $ V_{\text{top}} $, to the Higgs potential in Eq.~(\ref{Potential 1}).
The choice for these specific spurions will be clear when discussing the specific form of the potential.

Upon hyper-fermion condensation, the couplings in Eq.~\eqref{eq: top PC operators} generate linear mixing of the top spinors with
the baryons (i.e. spin-1/2 resonances) associated to the operators made of hyper-fermions. We will estimate the effect of such
couplings by constructing operators in terms of the spurions, following Ref.~\cite{Alanne:2018wtp}.
The operator generating the top mass reads:
\begin{equation}
\begin{aligned} \label{eq: Ltop PC}\mathcal{L}_{\rm top}&=\frac{C_t y_{L}y_{R}f}{4\pi}(Q_Lt_R^c)^\dagger \text{Tr}[P_Q^\alpha\Sigma P_t]+\text{ h.c.}, \end{aligned} 
\end{equation} 
where $C_t$ is an $\mathcal{O}(1)$ form factor, while $y_{L/R}$ are related to the couplings $\widetilde{y}_{L/R}$ via the anomalous dimensions of the fermionic operators, and are expected to be $\mathcal{O} (1)$ for the top.

The choice of spurions we did here is motivated by the fact that, at leading order in the chiral expansion, only the right-handed top spurion contributes to the potential:
\begin{equation}
\begin{aligned} \label{eq: V top p2} V_{\rm top,p^2}=&\frac{C_R f^4}{4\pi}y_R^2 \text{Tr}[P_t^\dagger\Sigma P_t^T \Sigma^\dagger],
\end{aligned} \end{equation} 
while a $ y_L^2 $ potential term is not allowed if the symmetric representation is chosen for the left-handed top. 
We further fix the components of $P_t$ as follows:
\begin{equation}
\begin{aligned} & A_t =1, \quad B_t = -\frac{1}{\sqrt{2}}, \quad C_t = \alpha e^{i\delta}, 
\end{aligned} \end{equation}
such that the leading order operator does not depend on the misalignment angle $\theta$. 
With this choice, the misalignment angle is determined by next-to-leading order terms in the top, which are small enough to keep the Higgs mass naturally close to the experimental value~\cite{Alanne:2018wtp}.
With this assumption, from Eq.~(\ref{eq: Ltop PC}) the mass and the HY coupling constant of the top quark are 
\begin{equation}
\begin{aligned} \label{eq: top mass and Yukawa} m_t = \frac{C_t v_{\rm EW}}{8\sqrt{2}\pi}y_L y_R, \quad\quad y_{ht\overline{t}}=\frac{m_t}{v_{\rm EW}}c_\theta, 
\end{aligned} \end{equation} 
where the top-HY coupling is SM-like for $ c_\theta \sim 1 $.\\


\textbf{NLO potential and the Higgs mass:} 
At next-to-leading order (NLO), may operators contribute to the potential~\cite{Alanne:2018wtp}.
In order to give a simple result, we will assume here that only double-trace operators are relevant, given by the following two terms:
\begin{equation}
\begin{aligned} \label{eq: V top p4} V_{\rm top,p^4}=&\frac{C_{\rm LL}f^4}{(4\pi)^2}y_L^4 \sum_{\alpha,\beta=1}^2 \text{Tr}[P_Q^\alpha \Sigma^\dagger P_Q^\beta \Sigma^\dagger]\text{Tr}[P_{Q,\alpha}^\dagger \Sigma P_{Q,\beta}^\dagger \Sigma]+ \\&\frac{C_{\rm LR}f^4}{(4\pi)^2}y_L^2 y_R^2 \sum_{\alpha=1}^2 \text{Tr}[P_Q^\alpha \Sigma^\dagger P_t^T]\text{Tr}[\Sigma P_{Q,\alpha}^\dagger P_t^\dagger] \\ =& \frac{f^4}{(4\pi)^2}\left(C_{\rm LL}y_L^4 s_\theta^4+C_{\rm LR}y_L^2 y_R^2 s_\theta^2 + \dots\right), 
\end{aligned} \end{equation} 
which contribute to the vacuum misalignment, because these terms depend on $ \theta $. Now, we have introduced the LO and NLO potential terms to the top potential $ V_{\rm top} = V_{\rm top,p^2}+V_{\rm top,p^4}+\dots $ in Eq.~(\ref{Potential 1}). 

By minimizing the potential in Eq.~(\ref{Potential 1}), we can fix the hyper-fermion mass term as a function of the misalignment angle as follows: 
\begin{equation}
\begin{aligned}  \label{eq: m12 expression}
Z \overline{m}=-\frac{\left(8 \pi^2 C_g \widetilde{g}^2 + 2C_{\rm LL}y_L^4 s_\theta^2+C_{\rm LR}y_L^2 y_R^2\right)f c_\theta}{64 \pi^3},
\end{aligned}
\end{equation} 
where $ \overline{m} \equiv m_1 +m_2 $ and $ \widetilde{g}^2\equiv 3g_L^2 + g_Y^2 $. Typically, $ C_{\rm LR}<0 $ while the other form factors $ C_{\rm t,g,R,LL}>0 $ in order to stabilize the vacuum for $\theta \neq 0$. From the total effective potential in Eq.~(\ref{Potential 1}) and the above vacuum misalignment condition, the physical SM Higgs, $ h $,  obtains a mass: 
\begin{equation}
\begin{aligned} \label{eq: Higgs mass expression}  
m_h^2=& \frac{v_{\rm EW}^2}{512\pi^2}\Big[(1+3c_{2\theta})C_{\rm LL}y_L^4-C_{\rm LR}y_L^2y_R^2-8\pi^2 C_g \widetilde{g}^2\Big] \,.
\end{aligned}
\end{equation}

The other pNGBs, including the components of the inert doublet $\eta$, will also receive a mass from the leading order operators, proportional to $C_R$.

\subsection{Lattice results and the Higgs mass tuning}

While the main focus of this work is on the generation of neutrino masses, which will be discussed in detail in the next section, we first provide a critical discussion on the value of the Higgs mass, which is one of the main concerns in the construction of realistic CH models. Additionally, we used models with a gauge-fermion underlying description as templates for our study. The results we derive here are based on an effective field theory approach, thus they apply to any model with the global symmetries we consider. Nevertheless, we comment here on the role lattice calculations have in validating such models, in view of the Higgs mass value.

In the previous section, we estimated the value of the Higgs mass via effective operators, hence introducing a dependence on unknown low energy constants. In the literature, the potential is also computed via loops of the resonances of the composite sector \cite{Contino:2011np,Contino:2015mha}, namely the heavy fermions that mix with the top in PC and the spin-1 resonances that mix with the $W$ and $Z$ EW bosons. To critically understand the main results, we first introduce a generic estimate for the Higgs mass, which reads:
\begin{equation}
m_h^2 \approx \frac{1}{(4 \pi)^2} y_{\rm SM}^2 M_{\rm res}^2 \sin^2 \theta\,.
\end{equation}
Here, the pre-factor derives from the loop, $y_{\rm SM}$ is a generic SM coupling (for instance the top Yukawa) and $M_{\rm res}^2$ is the mass of a generic composite resonance. The factor of $\sin^2 \theta = v_{\rm EW}^2/f^2$ can be understood in terms of symmetries \cite{Cacciapaglia:2020kgq}: for $\theta \neq 0$, the spectrum contains three massless Goldstones (the longitudinal $W$ and $Z$ polarizations); in the continuum limit $\theta \to 0$ (where, therefore, $v_{\rm EW} \to 0$) the EW symmetry is restored and the three massless Goldstones should be completed by a fourth degree of freedom in order to reconstruct a complete EW multiplet (the Higgs doublet). The new Goldstone at $\theta = 0$ is the Higgs boson, whose mass should vanish for  vanishing $\theta$. Using $M_{\rm res} = g^\ast f$, where $g^\ast \approx \mathcal{O} (1)$ is a coupling of the strong sector, and the relation between $v_{\rm EW}$ and the condensation scale $f$, the Higgs mass reads
\begin{equation}
m_h^2 \approx \frac{(g^\ast)^2}{(4 \pi)^2} y_{\rm SM}^2 v_{\rm EW}^2\,,
\end{equation}
in agreement with Eq.~\eqref{eq: Higgs mass expression}. This value is naturally around the EW scale, as found experimentally.

A more quantitative relation between the Higgs mass and the resonance masses can be computed via loops of the resonances, as done in the majority of the CH literature. In particular, explicit computations found that the mass of the top partners needs to be fairly light, $M_{\rm res} \lesssim 1.5$~TeV~\cite{Matsedonskyi:2012ym,Redi:2012ha}. This result, however, is based on some important assumptions, as discussed in Ref.~\cite{Contino:2011np}. In particular, the loops are only calculable if the resonances can be safely described in the effective field theory framework: one key requirement is that the masses are at most of order $f$ and the couplings ($g^\ast$) are in the perturbative regime (partial UV completion criterion). Hence, for masses $M_{\rm res} \gg f \approx 1$~TeV, the loop calculation cannot be trusted, and the Higgs mass can only be expressed in terms of operators, as we described in the previous section (for a complete basis, see Refs.~\cite{Alanne:2018wtp,Golterman:2017vdj}). Hence, for heavy top partners, the quantitative relation between the Higgs mass and the resonance masses cannot be trusted. Instead, the precise value of the Higgs mass is encoded in low energy constants, which can be computed on the lattice once an underlying theory is defined.

Lattice collaborations have started exploring the models we used as a template for our discussion, so it is interesting to summarize here the main results. A model based on the confining gauge symmetry $\SP(4)_{\rm TC}$ has been studied~\cite{Bennett:2017kga,Lee:2019pwp,Bennett:2019cxd}, with fermions in the fundamental representation and the top partners containing fermions in the anti-symmetric one. This is one of the model proposed in Refs.~\cite{Barnard:2013zea,Ferretti:2013kya}, which features the global symmetry pattern we need for this work. However, the current results cannot be used to exclude the model as a candidate for a CH model with PC, because the spectrum for the top partners is not available yet, as the fermions in the anti-symmetric representation are not fully dynamical in the most recent simulations \cite{Bennett:2019cxd}.
A study of models based on $\SU(4)_{\rm TC}$ is at a more advanced status \cite{Ayyar:2017qdf,Ayyar:2018zuk,Ayyar:2019exp,Svetitsky:2019hij}, however targeting a different class of cosets. 
Results with fully dynamical fermions show that the spin-1/2 resonances, candidates to be top partners, have masses much larger than $f$ \cite{Ayyar:2018zuk}. This is also confirmed by holographic techniques \cite{Erdmenger:2020flu}. However, as explained above, the loop calculations for the Higgs potential are not trustable in this limit, and we would need the values of the appropriate low energy constants, which have not been computed yet. So far, only results available concerns the operator generated by gauge loops \cite{Ayyar:2019exp}, which has a minor impact on the value of the Higgs mass.

In summary, within the current  knowledge of the strong dynamics needed to generate the model under study in this work, it is feasible that the value of the Higgs mass may match the experimentally observed one. A mild tuning in the parameters may be needed, without any consequences that are excluded by other experimental observations.

\subsection{Neutrino Yukawa coupling constants}\label{IIIB}

Having discussed the top mass generation, we are now ready to introduce the couplings responsible for generating masses for neutrinos. 
The other charged fermions, lighter than the top, have minor effect on the Higgs physics, and do not spoil the scotogenic mechanism as long as they decouple from the inert Higgs doublet. Here we will consider three different approaches. \\

\textbf{(i) ETC-type four-fermion operators:} Firstly we add four-fermion operators like those in Ref.~\cite{Kaplan:1983sm}, bilinear in the hyper-fermions, which could arise from the exchange of heavy scalar multiplets or from heavy vectors as in ETC models~\cite{Appelquist:2002me,Appelquist:2003hn}. These operators are the analogous to the four-fermion interactions used in Ref.~\cite{Cacciapaglia:2014uja} and, in our model, can be written as
\beq \label{four fermion operators neutrinos}
- \frac{y_\nu^{ij}}{\Lambda_\nu^2}(L_{L,i} N_{R,j}^{c})_\alpha^\dag(\Psi_2^\alpha \psi_3) + \mbox{h.c.}, \phantom{0000}
\eeq 
where it is assumed the Yukawa couplings $ y_\nu^{ij} \sim \mathcal{O}(1) $, and $ \Lambda_{\nu}>\Lambda_{\rm HC} $ is the energy scale where such interactions are generated. We will leave this part unspecified. The hyper-fermion bilinear $ \Psi_{2}^\alpha \psi_3 $ transforms as the Higgs doublet $ \eta $. 

Upon condensation, the above coupling generates the operator: 
\beq \label{condense Yukawa operators}
- h^{ij} f (L_{L,i} N_{R,j}^c)_\alpha^\dag \text{Tr}[P_{\eta}^\alpha \Sigma] + \mbox{h.c.}\phantom{00}
\eeq 
with the couplings $ h^{ij} \equiv  4\pi N A (\Lambda_{\text{HC}}/\Lambda_\nu)^2 y_{\nu}^{ij} $~\cite{Hill:2002ap} where $ A $ is an integration constant arising from the condensation 
and $ N $ is the number of hyper-colors. Here $P_{\eta}^\alpha$ are projectors that extract the $\eta$ component of the pNGB matrix. Note also that the suppression by $\Lambda_\nu$ could be softened if the fermion bilinear has a sizeable anomalous dimension in the walking window above $\Lambda_{\rm HC}.$\\

\textbf{(ii) Fermion fundamental partial compositeness:} Fermion masses in CH models can also be generated via fundamental Yukawa couplings involving scalars charged under the confining HC gauge interactions~\cite{Sannino:2016sfx,Cacciapaglia:2017cdi}. To implement this mechanism for neutrinos, we need to extend our underlying model with three $ \mathbb{Z}_2\mbox{--odd} $ techni-scalars $ S_{\nu,i} $ with hypercharge $ Y=+1/2 $, and transforming in the same representation as $\Psi$ under $ G_{\rm HC} $. At the fundamental Lagrangian level, the new Yukawa couplings with the neutrino fields read 
\beq  \label{yukawa couplings ii}
y_{\nu}^{ij} \epsilon_{\alpha\beta} L_{L,i}^\beta S_{\nu,j} \epsilon_{\rm HC} \Psi_2^\alpha - y_{N}^{ij} S_{\nu,i}^* N_{R,j}^c \psi_4 +\rm h.c.,
\eeq 
where $ \alpha, \beta  $ are $ \SU(2)_{\rm L} $ indices, and the coupling constants $ y_{\nu,N} $ are $ 3\times 3 $ matrices in flavour space. The techni-scalars $S_{\nu,i}$ inherit the $ \mathbb{Z}_2\mbox{--odd} $ parity via these interactions.

A general operator analysis of the low energy effective description for this class of theories has been presented in Ref.~\cite{Cacciapaglia:2017cdi}. After the hyper-fermions and -scalars condense, the content of the low energy theory depends crucially on the mass of the techni-scalars: if $M_S \gg \Lambda_{\rm HC}$, then the scalars can be integrated out before the condensation and generate effective four-fermion interactions like those in Eq.~\eqref{four fermion operators neutrinos}; if $M_S \ll \Lambda_{\rm HC}$, at low energies spin-1/2 resonances made of one scalar and one fermion will be present, which mix with the neutrino fields, thus generating fermion partial compositeness. This scenario can, therefore, be considered as an intermediate case between ETC-like interactions and partial compositeness.

In the former case, after integrating out the scalars, the low energy theory will contain the coupling below:  
\beq \label{eff yukawa couplings ii}
-h^{ij}\nu_{L,i}  N_{R,j}^c \eta^0 +\rm h.c.,
\eeq
where $ h^{ij}\equiv 4\pi N C_{\rm Yuk} \Lambda_{\rm HC}^2 \sum_{k,l}(1/M_S^2)_{kl}y_{\nu}^{ik} y_{N}^{lj} $, $ C_{\rm Yuk} $ is an $ \mathcal{O}(1) $ non-perturbative coefficient, and $ (M_S)_{ij} $ is the mixing mass matrix of the techni-scalars $ S_{\nu,i} $. This approach can easily be extended to include the masses and mixing of the other fermions as shown in Ref.~\cite{Cacciapaglia:2017cdi}, and also provide mass for the top via partial compositeness (however the operators mixing with the top will necessarily transform as the fundamental of $\SO(6)$, thus giving different contributions to the potential that discussed in the previous section).  \\

\textbf{(iii) Fermion partial compositeness:} Finally, as in Ref.~\cite{Frigerio:2018uwx}, we can postulate that neutrinos -- like the top quark -- weakly mix with composite operators $ O^{\psi}_a $ at some UV scale, $ \Lambda_{\rm UV} $, in the following manner: 
\beq \label{yukawa couplings iii}
\lambda_{ai}^L L_{L,i}^\alpha \overline{O}^{L}_{a,\alpha} + \lambda_{ai}^N N_{R,i} \overline{O}^{N}_a + \rm h.c.,
\eeq 
where $ \lambda^{L,N}_{ai} $ are coupling constants, $ \alpha $ is the $ \SU(2)_{\rm L} $ index of the fermion doublets, $ i=1,2,3 $ is the generation index of the neutrinos, and $ a=1,2,3 $ is a flavour index of the new dynamics. At the condensation scale $ \Lambda_{\rm HC} $, the operators $ O^{\psi}_a $  are matched to massive spin-1/2 resonances, which pass the couplings to the pNGBs, including the $ \mathbb{Z}_2 $--odd doublet $ \eta $, to the neutrino fields. Schematically, we will assume that the spin-1/2 masses are equal to $ m_*^a $, and their couplings are controlled by a single parameter $ g_* \in [1,4\pi] $. 

Above $\Lambda_{\rm HC}$, the theory flows to a conformal phase (walking), which is crucial in this class of model in allowing a large enough top Yukawa coupling via partial compositeness, while keeping the heavy scales $\Lambda_{\rm UV}$ large enough to avoid flavour bounds. We will therefore review how the anomalous dimension of the operators influence the size of the couplings in Eq.~\eqref{yukawa couplings iii}.
At leading order in small conformal field theory perturbations between the scales $ \Lambda_{\rm UV} $ and $ m_*^a $, the renormalization group evolution of the couplings $ \lambda_{ai}^\psi $ is given by~\cite{Frigerio:2018uwx}
\beq
\mu \frac{d}{d\mu}\lambda_{ai}^\psi = (\Delta_a^\psi-5/2)\lambda_{ai}^\psi + \mathcal{O}(\lambda^3),
\eeq 
where $ \Delta_a^\psi \equiv \Delta [O_a^\psi]$ is the definite scaling dimensions of the operators $ O_a^\psi $. The solution of this RG evolution for $ m_* \ll \Lambda_{\rm UV} $ is 
\beq \label{lambdaofmstar}
\lambda_{ai}^\psi (m_*)= \left( \frac{m_*}{\Lambda_{\rm UV}} \right)^{\Delta_a^\psi -5/2} \  \lambda_{ai}^\psi  (\Lambda_{\rm UV})\,.
\eeq 
Thus, for $ \Delta_a^\psi > 5/2 $ the coupling at low energies is strongly suppressed (corresponding to an irrelevant operator), while for $ \Delta_a^\psi \leq 5/2 $ it tends to grow at lower energies and may reach a non-trivial infra-red fixed point. For the top, one clearly needs large dimensions in order to achieve large-enough HY couplings.  For the neutrino couplings in Eq.~\eqref{yukawa couplings iii}, both enhanced and suppressed couplings may be viable, as they will ultimately be connected to the masses of the right-handed neutrinos, $M_i$. In the phenomenological analysis, we will therefore allow the neutrino partial compositeness mixing terms to span a wide range of values, from very suppressed to sizeable.

 
 Effective Yukawa couplings are generated below $m_*^a$, where the spin-1/2 states are integrated out.
The hierarchy of $ \lambda_{ai}^\psi (m_*^a) $ will thus translate into a hierarchy of the Yukawa couplings $ h^{ij} $ by redefining the elementary fermions $ L_{L,i} $ and $ N_{R,i} $ via unitrary rotations, \begin{equation}
\begin{aligned} \label{eff yukawa couplings iii} h^{ij}&=g_* (\epsilon_{ai}^L)^*\epsilon_{bj}^N c_{ab}= g_*  \begin{pmatrix}  \epsilon_1^L \epsilon_1^N c_{11} & \epsilon_1^L \epsilon_2^N c_{12} & \epsilon_1^L \epsilon_3^N c_{13} \\  \epsilon_2^L\epsilon_1^N c_{21}&  \epsilon_2^L \epsilon_2^N c_{22} &   \epsilon_2^L \epsilon_3^N c_{23} \\\epsilon_3^L\epsilon_1^N c_{31} &\epsilon_3^L \epsilon_2^N c_{32}& \epsilon_3^L \epsilon_3^N c_{33}\end{pmatrix}, \end{aligned}  \end{equation}
where $ g_* $ is the strong-sector low-energy coupling, $ \epsilon_i^\psi \equiv \lambda^\psi_{ii}/g_* $, and $ c_{ab} $ are model-dependent parameters of order unity from the strong dynamics. Therefore, the parameters $ \epsilon_i^\psi $ inherit the suppression/enhancement from Eq.~\eqref{lambdaofmstar}. These parameters encode a measure of ``compositeness'' of the SM fermions at scales of order $ m_*^a $, and without loss of generality they are real, positive and normalised to one in the limit of fully composite fermions. 

To generate the operators $ O_{a,\alpha}^{L} $ and $ O_a^{N} $ in Eq.~(\ref{yukawa couplings iii}) in our underlying theory, we will introduce three new hyper-fermions, $ \chi_{\nu,a} $, transforming in the two-index anti-symmetric representation of $ G_{\rm HC}=\SP(2N)_{\rm HC} $, analogous to the $\chi_t$ fermions added for the top. However, the $ \chi_{\nu,a} $ are singlets under the SM gauge interactions.
For concreteness, we will consider, in analogy to the top in Eq.~(\ref{eq: top PC operators}), the following four-fermion interactions generated at $\Lambda_{\rm UV}$:
\beq \label{PCeffeciveops}
\frac{\widetilde{y}_\nu^{ia}}{\Lambda_\nu^2} L^{\alpha\dagger}_{L,i} (\Psi^\dagger P_L^\alpha \Psi^*\chi_{\nu,a}^\dagger) + \frac{\widetilde{y}_N^{ia}}{\Lambda_\nu^2}N_{R,i}^\dagger (\Psi^\dagger P_N \Psi \chi_{\nu,a}^\dagger)+\rm h.c. \ \ \ \ \ \
\eeq 
For the left-handed neutrinos, the spurions can take either the symmetric (with $ + $) or the anti-symmetric ($ - $) representation of the chiral symmetry $ \SU(6) $:
\begin{equation}\begin{aligned} \label{spurions left} 
P_{L,ij}^1=\frac{1}{\sqrt{2}}(\delta_{i3}\delta_{j5}\pm\delta_{i5}\delta_{j3}), \ P_{L,ij}^2=\frac{1}{\sqrt{2}}(\delta_{i4}\delta_{j5}\pm\delta_{i5}\delta_{j4}),  \end{aligned}  \end{equation}
while for the right-handed neutrinos the spurion transforms as the adjoint representation of $ \SU(6) $:
 \begin{equation}
\begin{aligned} \label{spurions right}  P_N=A_N P_N^1 +B_N P_N^2 + C_N P_N^3,  
\end{aligned}  \end{equation} 
where 
\begin{equation}
\begin{aligned}  & P_N^1=\frac{1}{\sqrt{2}}\text{Diag}(0,0,1,-1,0,0),\nonumber \\ & P_N^2=\frac{1}{2}\text{Diag}(0,0,-1,-1,1,1),\nonumber \\ \nonumber &P_N^3=\frac{1}{2\sqrt{3}}\text{Diag}(-2,-2,1,1,1,1) .
\end{aligned} \end{equation} 
The naive scaling dimension of these operators is $\Delta = 9/2 > 5/2$, thus the couplings will be suppressed by $(m_*^a/\Lambda_{\rm UV})^2$, unless a large anomalous dimension is generated in the walking window.

Below $\Lambda_{\rm HC}$, the two operators in Eq.~\eqref{PCeffeciveops} are matched to composite resonances with the same quantum numbers, respectively, of the lepton doublets and of the right-handed neutrinos, which couple to each other via the inert Higgs doublet $\eta$. We will assume, for simplicity, that both resonances receive a mass $m_* \equiv m_*^a$ from the strong dynamics, where the mass of the composite singlet can be of Majorana type.
Thus, below $m_*$, the mixing of the SM fields with the composite resonances generates the effective Yukawa couplings for $ M_i < m_* $
\begin{equation}\begin{aligned} 
\label{eff yukawa couplings iii specific} h^{ij}&=  \frac{A_N-\sqrt{2}B_N}{16\sqrt{2}\pi}  g_* \sum_{a,b=1}^3 (\epsilon_{ia}^L)^* c_{ab} \epsilon_{bj}^N\\ &\equiv \frac{3}{16\sqrt{2}\pi}  g_* \sum_{a,b=1}^3 (\epsilon_{ia}^L)^* c_{ab} \epsilon_{bj}^N.
\end{aligned}  \end{equation} 
Only the combination $A_N-\sqrt{2}B_N$ enters the neutrino Yukawa, thus to simplify the equations in the last line we have fixed $A_N = 1$ and $B_N = - \sqrt{2}$. We will use this redefinition in the numerical calculations in Section~\ref{Num results}. 

In the case with right-handed neutrino masses $ M_i\gg m_* $, explained in Appendix~\ref{sec: Appendix A}, we obtain the Yukawa couplings in Eq.~(\ref{eff yukawa couplings iii specific heavy appendix}), \begin{equation}
\begin{aligned} \label{eff yukawa couplings iii specific 2}
h^{ai}= \frac{3}{16\sqrt{2}\pi} g_* \sum_{k,b,c=1}^3 \frac{ m_\ast^a}{M_k}\epsilon_{ak}^N \epsilon_{kc}^N c_{cb} (\epsilon_{bi}^L)^*.
\end{aligned}
 \end{equation} These couplings are the modified HY couplings of $ h^{ij} $ in Eq.~(\ref{eff yukawa couplings iii specific}), where the right-handed neutrinos, $ N_{R,i} $, are integrated out and replaced by the right-handed composite neutrino partners, $ O_{N,a} $.

\subsection{The masses and mixing of the neutrinos}

At the leading order, the model only generates a mass for the top quark in Eq.~(\ref{eq: top mass and Yukawa}) (and similarly for the other charged SM fermions), while neutrinos remain massless. For the top, we have used a specific set of PC operators, but the top mass could also be generated by the other two mechanisms used for neutrinos. 

Furthermore, besides the $\mathbb{Z}_2$--odd doublet, the pNGB spectrum contains other odd states (i.e. a triplet and a singlet) that mix with the doublet $\eta$ components. In particular, the mixing mass matrices $ M_{R}^2 $ and $ M_{I}^2 $ in the bases $ (\rm  Re\eta^0, \Delta^0) $ and $ (\rm Im\eta^0, \varphi^0) $ are generated by the potential, which are given by \begin{equation}
\begin{aligned} 
M_R^2 = \begin{pmatrix} M_{R,11}^2 & M_{R,12}^2 \\ M_{R,21}^2 & M_{R,22}^2\end{pmatrix}, \quad \quad M_I^2 = M_R^2 + C_gf^2 g_L^2 \begin{pmatrix} s_{\theta/2}^2 & 0 \\ 0 & -c_{\theta/2}^2\end{pmatrix}. 
\end{aligned}
 \end{equation} with \begin{equation*}
\begin{aligned} 
M_{R,11}^2=&\frac{f}{512\pi^2}\Big[f\big(16\pi^2 C_g\widetilde{g}^2 +8\pi C_Ry_R^2-3C_{LL}y_L^4-2C_{LR}y_L^2y_R^2 \big)+512\pi^3 Z m_3  \\ &\phantom{\frac{f}{512\pi^2}\Big[}- C_{LL}fy_L^4c_{4\theta}+ 2f\big(8\pi^2 C_g \widetilde{g}^2 +2C_{LL}y_L^4+C_{LR}y_L^2y_R^2\big)c_{2\theta} \\& \phantom{\frac{f}{512\pi^2}\Big[}+16\pi\big(2\pi C_g f \widetilde{g}^2+C_R f y_R^2+ 16\pi^2 Z m_{12} \big)c_\theta\Big], \\ 
M_{R,22}^2=&M_{R,11}^2+\frac{f^2}{16\pi}\Big[8\pi C_g g_L^2+\big(2\pi C_g(g_L^2-g_Y^2)-C_Ry_R^2\big)c_\theta\Big], \\
M_{R,12}^2=&M_{R,21}^2=\frac{f^2 s_\theta}{128 \pi^2}\Big[4\pi C_Ry_R^2  -C_{LR}y_L^2y_R^2-2C_{LL}y_L^4 s_{\theta}^2\Big].
\end{aligned}
 \end{equation*} In the following, we will only need to consider the two lightest eigenstates, $ \rm Re\widetilde{\eta}^0 $ and $ \rm Im\widetilde{\eta}^0 $ consisting mostly of $ \rm Re\eta^0 $ and $ \rm Re\eta^0 $, with masses $ m_R $ and $ m_I $, respectively. The expressions above show that the mass difference between them is due to gauge corrections proportional to $C_g$, which are typically small, and suppressed by two powers of $s_{
\theta/2}$, hence their near-degeneracy.

For light right-handed neutrinos, neutrino masses are generated by the loop in Fig.~\ref{loopdiagram}, analog to the one in the traditional scotogenic model~\cite{Ma:2006km}. This situation can be achieved in all three mechanisms introduced above, as long as  $ M_i<\Lambda_{i} $ with $ \Lambda_{i} \equiv \Lambda_\nu$ for ETC, $ \Lambda_{i} \equiv M_S$ for FPC, and $ \Lambda_{i}\equiv m_\ast$ for the PC-type approach.
The loop results in the following expression~\cite{Ma:2006km}
 \begin{equation}
\begin{aligned} \label{top neutrino mass}
m_{\nu}^{ij}=&&\sum_{k=1}^{3} \frac{h^{ik}h^{kj}}{(4\pi)^2}M_k\Bigg[\frac{m_R^2}{m_R^2-M_k^2}\ln\left(\frac{m_R^2}{M_k^2}\right) -\frac{m_I^2}{m_I^2-M_k^2}\ln\left(\frac{m_I^2}{M_k^2}\right)\Bigg] ,
\end{aligned}
 \end{equation} 
 where we recap the expressions of the neutrino HY coupling constants:
 \begin{equation}
\begin{aligned}  \label{eq: Yukawa coupling ETC FPC PC}
\textbf{ETC: } &h^{ij}= 4\pi N A \left(\frac{\Lambda_{\rm HC}}{\Lambda_\nu}\right)^2 y_{\nu}^{ij}, \\ 
\textbf{FPC: } &h^{ij}= 4\pi N C_{\rm Yuk} \Lambda_{\rm HC}^2 \sum_{k,l} \left(\frac{1}{M_S^2}\right)_{kl} y_{\nu}^{ik} y_{N}^{lj}, \\
\textbf{PC: } &h^{ij}=\frac{3}{16\sqrt{2}\pi}  g_* \sum_{a,b} (\epsilon_{ai}^L)^*\epsilon_{bj}^N c_{ab},
\end{aligned}
 \end{equation}
 which are valid for $ \Lambda_i > \Lambda_{\rm HC} $. For simplicity, we assume in the following that $ (1/M_S^2)_{ij} $ is diagonal with identical diagonal elements, namely $ M_S $. The case with FPC in which the masses of the techni-scalars, $ M_S $, are smaller than $ \Lambda_{\rm HC} $ is obtained by removing $ (\Lambda_{\rm HC}/M_S)^2 $~\cite{Cacciapaglia:2017cdi}.

\begin{figure}[t!]
	\centering
	\includegraphics[width=0.45\textwidth]{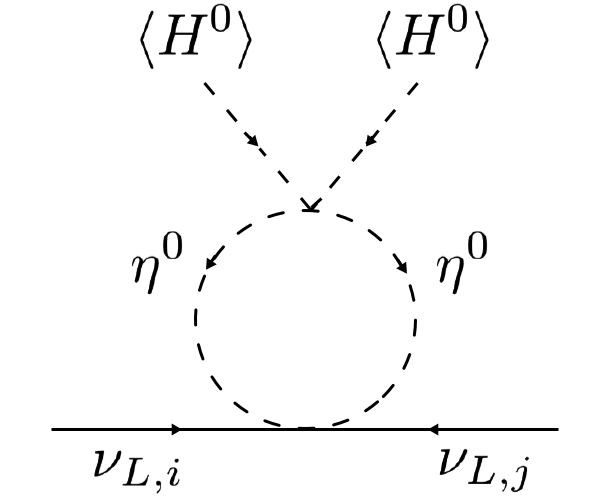}
	\caption{One-loop Majorana neutrino mass for heavy right-handed neutrinos in the composite scenarios, with ETC and FPC mechanisms.}
	\label{loopdiagram2}
\end{figure}
 
In the ETC and FPC cases, for $ M_i\gg \Lambda_{i} $, the right-handed neutrinos must be integrated out before the theory condenses. This will effectively generate a direct coupling of the left-handed neutrinos with two inert Higgs doublets, in a form similar to the Weinberg operator, shown in Eq.~(\ref{eq: Weinberg operator ETC}) and~(\ref{eq: Weinberg operator FPC}) in Appendix~\ref{sec: Appendix A} for the ETC and FPC case, respectively. Thus, the neutrino mass will be generated by the loop in Fig.~\ref{loopdiagram2}, and it can be expressed as
 \begin{equation}
\begin{aligned} \label{top neutrino mass Mi large}
m_{\nu}^{ij}= \sum_{k=1}^{3} \frac{h^{ik}h^{kj}}{(4\pi)^2 M_k}\Bigg[&\Lambda_{\text{HC}}\sqrt{m_R^2+\Lambda_{\text{HC}}^2}+ m_R^2 \ln\Bigg(\frac{m_R}{\Lambda_{\text{HC}}+\sqrt{m_R^2+\Lambda_{\text{HC}}^2}}\Bigg) \\& -\Lambda_{\text{HC}}\sqrt{m_I^2+\Lambda_{\text{HC}}^2}- m_I^2 \ln\Bigg(\frac{m_I}{\Lambda_{\text{HC}}+\sqrt{m_I^2+\Lambda_{\text{HC}}^2}}\Bigg) \Bigg], 
\end{aligned}
 \end{equation} which is calculated in Eqs.~(\ref{eq: loop calculation appendix}) and~(\ref{eq:neutrino mass matrix}) in Appendix~\ref{sec: Appendix A}. 

In the PC case, a similar expression as in Eq.~(\ref{top neutrino mass}) holds for $M_i \gg m_\ast^a$, where integrating out the elementary right-handed neutrinos simply generated a correction to the Majorana mass of the composte singlets. Thus, it suffices to replace $M_i \to m_\ast^a$ in the above expression given in Eq.~(\ref{top neutrino mass}). In other words, for the PC case, $m_\ast^a$ generated by the strong dynamics acts as a cap for the mass of the state that plays the role of the right-handed neutrinos. From the results in Eq.~(\ref{top neutrino mass PC appendix}) in Appendix~\ref{sec: Appendix A}, the loop results in Eq.~(\ref{top neutrino mass}) are now rewritten as \begin{equation}
\begin{aligned} \label{top neutrino mass PC}
m_{\nu}^{ij}=&&\sum_{a=1}^{3} \frac{h^{ia}h^{aj}}{(4\pi)^2}m_\ast^a\Bigg[\frac{m_R^2}{m_R^2-(m_\ast^a)^2}\ln\left(\frac{m_R^2}{(m_\ast^a)^2}\right) -\frac{m_I^2}{m_I^2-(m_\ast^a)^2}\ln\left(\frac{m_I^2}{(m_\ast^a)^2}\right)\Bigg],
\end{aligned}
 \end{equation} where the modified Yukawa couplings $ h^{ai} $ are given in Eq.~(\ref{eff yukawa couplings iii specific 2}).

Thus, for $ M_i \gg \Lambda_i $, the ETC and FPC approaches of neutrino mass generation may be distinguished from the PC approach by considering the different form of the neutrino mass matrices in Eqs.~(\ref{top neutrino mass Mi large}) and~(\ref{top neutrino mass PC}). Furthermore, in Ref.~\cite{Alanne:2016rpe}, it was shown that the composite pNGB $ \chi $ in Table~\ref{tab:su6sp6} and $ \chi' $ corresponding to a quantum anomalous $ \rm U(1) $ may be used to disentangle the three different fermion (possibly neutrino) mass mechanisms we consider here. In Refs.~\cite{Belyaev:2016ftv,Cacciapaglia:2017iws}, these composite states have been studied in a PC scenario with both an EW sector and a QCD-colored sector. Finally, the heavy particles (the ETC gauge bosons, ETC scalars, techni-scalars $ S_{\nu,i} $, or the composite neutrino partners) 
providing the different operators in Eq.~(\ref{four fermion operators neutrinos}) for ETC, Eq.~(\ref{yukawa couplings ii}) for FPC, and Eq.~(\ref{yukawa couplings iii}) for PC may be observable in future collider experiments.

The above neutrino mass matrices can be diagonalized as \begin{equation}
\begin{aligned}
	m_\nu^{\rm Diag} &= U_{\rm PMNS}^T m_\nu U_{\rm PMNS}={\rm Diag}(m_{\nu_1},m_{\nu_2},m_{\nu_3}),  
\end{aligned}
 \end{equation} where $ m_{\nu_{i}} $ with $ i=1,2,3 $ are the left-handed neutrino masses. The matrix $ U_{\rm PMNS}=U U_m   $ is the PMNS matrix where $ U_m =\rm Diag(1,e^{i\phi_1/2},e^{i\phi_2/2}) $ encoding the Majorana phases and the matrix $ U $ is parametrized as \beq
\begin{pmatrix}c_{12}c_{13}&s_{12}c_{13}&s_{13}e^{-i\delta}\\-s_{12}c_{23}-c_{12}s_{23}s_{13}e^{i\delta} & c_{12}c_{23}-s_{12}s_{23}s_{13}e^{i\delta}  & s_{23}c_{13} \\ s_{12}s_{23}-c_{12}c_{23}s_{13}e^{i\delta}& -c_{12}s_{23}-s_{12}c_{23}s_{13}e^{i\delta}&c_{23}c_{13}\end{pmatrix} \nonumber
\eeq with the Dirac phase $ \delta $. In this paper, we assume the Majorana phases are vanishing ($ \phi_{1,2}=0 $), but it is possible to add them without significant changes of our conclusions. 

In the following section, we fit to the best-fit experimental values for the mass-squared differences ($ \Delta m_{ij}\equiv m_{\nu_i}^2-m_{\nu_j}^2 $) and mixing angles ($ s_{ij}\equiv \sin\theta_{ij} $), which are~\cite{Esteban:2018azc} \beq \label{best-fit experimental values}
	&& \Delta m_{31}^2 = 2.528^{+0.029}_{-0.031}\times 10^{-3}\text{ eV}^2,  \\  
	&&\Delta m_{21}^2 = 7.39^{+0.21}_{-0.20}\times 10^{-5}\text{ eV}^2, \nonumber \\
	&&s_{12}^2=0.310^{+0.013}_{-0.012}, \ s_{13}^2=0.02237^{+0.00066}_{-0.00065}, \ s_{23}^2=0.563^{+0.018}_{-0.024}, \nonumber
\eeq for normal hierarchy (NH), i.e. $ m_{\nu_1} < m_{\nu_2} < m_{\nu_3} $, and  \beq \label{best-fit experimental values IH}
	&& \Delta m_{32}^2 = -2.510^{+0.030}_{-0.031}\times 10^{-3}\text{ eV}^2,  \\  
	&&\Delta m_{21}^2 = 7.39^{+0.21}_{-0.20}\times 10^{-5}\text{ eV}^2, \nonumber \\
	&&s_{12}^2=0.310^{+0.013}_{-0.012}, \ s_{13}^2=0.02259^{+0.00065}_{-0.00065}, \ s_{23}^2=0.565^{+0.017}_{-0.022}, \nonumber
\eeq for inverted hierarchy (IH), i.e. $ m_{\nu_3} < m_{\nu_1} < m_{\nu_2} $. 

So far, an upper bound exists on the sum of the neutrino masses coming from cosmology. The most reliable bound is from the Planck collaboration~\cite{Ade:2015xua}, \beq
	\sum_i m_{\nu_i} \lesssim 0.23 \rm \ eV.
\eeq By using the measurements of $ \Delta m_{ij}^2 $ and the upper bound of the sum of the neutrino masses, we obtain an upper bound on the mass of the lightest neutrino ($ i=1 $ for NH and $ i=3 $ for IH): \beq
	 m_{\nu_i}^{\rm lightest} \lesssim 0.07 \rm \ eV.
\eeq 

\begin{figure}[t!]
	\centering
	\includegraphics[width=0.7\textwidth]{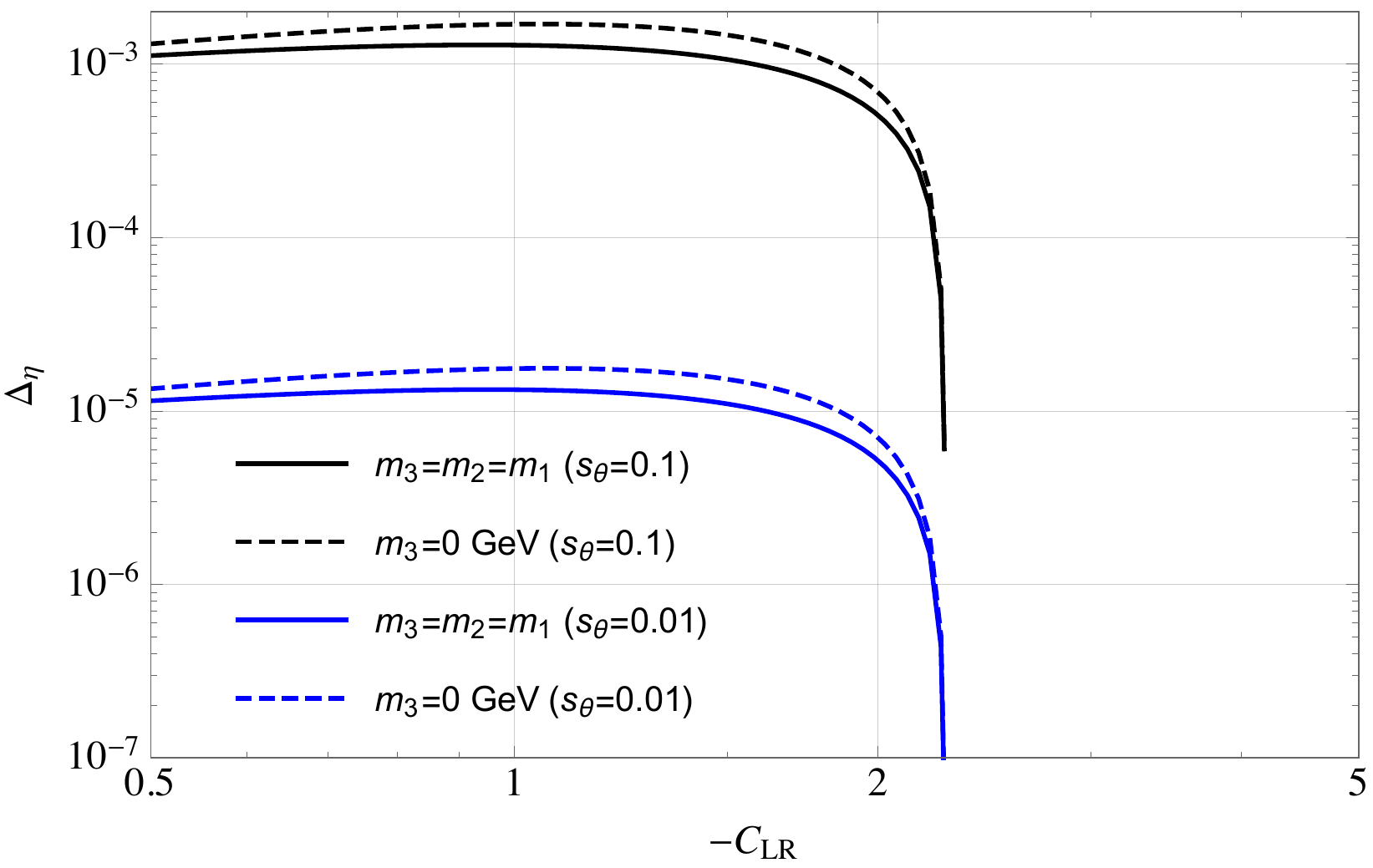}
	\caption{Level of degeneracy $\Delta_\eta \equiv (m_R-m_I)/m_I  $ of the scalars in the neutrino mass loop as function of $- C_{\rm LR}$. The curves correspond to different values of $s_\theta = 0.1$ and $0.01$, and for different choices of the hyper-fermion masses: democratic $m_1 = m_2 = m_3$, and $m_3 = 0$. The degeneracy shows a mild dependence on the hyper-fermion masses.}
	\label{ReImEta0 masses}
\end{figure}

\section{Numerical results}
\label{Num results}

We are now ready to numerically analyze the Yukawa couplings, $ h^{ij} $ in Eqs.~(\ref{top neutrino mass}) and~(\ref{top neutrino mass Mi large}) or $ h^{ai} $ in Eq.~(\ref{top neutrino mass PC}), generated in the concrete composite 2HDM we use here. We will fit the theoretical expressions to the best-fit experimental values in Eq.~(\ref{best-fit experimental values}) for NH and Eq.~(\ref{best-fit experimental values IH}) for IH. 
The expressions for the EW VEV in Eq.~(\ref{WZ masses and SM VEV}), the top mass in Eq.~(\ref{eq: top mass and Yukawa}), and the SM Higgs mass in Eq.~(\ref{eq: Higgs mass expression}) can be fixed to their observed values~\cite{Tanabashi:2018oca}, and $Z \overline{m} $ can be eliminated by the vacuum misalignment condition for $ \theta $ in Eq.~(\ref{eq: m12 expression}) from the minimization of the effective potential. Here, we have recollected these expressions: \begin{equation}
\begin{aligned} 
&v_{\rm EW}=2\sqrt{2}f s_\theta \approx 246~\mathrm{GeV}, \\ & m_h^2= \frac{1}{512\pi^2}\Big[(1+3c_{2\theta})C_{\rm LL}y_L^4-C_{\rm LR}y_L^2y_R^2-8\pi^2 C_g \widetilde{g}^2\Big]v_{\rm EW}^2\approx (125~\mathrm{GeV})^2, \\ & m_t = \frac{C_t v}{8\sqrt{2}\pi}y_L y_R\approx 173~\mathrm{GeV},\quad
 \\ & Z \overline{m} =-\frac{\left(8 \pi^2 C_g \widetilde{g}^2 + 2C_{\rm LL}y_L^4 s_\theta^2+C_{\rm LR}y_L^2 y_R^2\right)f c_\theta}{64 \pi^3},
\end{aligned}
 \end{equation} 
 which allow us to eliminate four free parameters (for instance, $Z \overline{m}$, $y_L$, $y_R$ and $f$). 
Without loss of generality, we can reabsorb the two form-factors $C_{\rm LL}$ and $C_{\rm R}$ in the $y_{L/R}$ (an elimination of them from the equations by setting $C_{\rm LL} = C_{\rm R} = 1$).  
Finally, to simplify the analysis, we fix to unity the remaining form factors, i.e. $ C_{t}=C_{g}=-C_{\rm LR}=1 $, and fix $ s_\theta=0.1$ ($f = 870$~GeV), 
 so that the only free parameters are the HY coupling constants, $ h^{ij} $ or $ h^{ai} $. The latter choice should be considered as a benchmark point for the model.

\begin{figure}[t!]
	\centering
	\includegraphics[width=0.49\textwidth]{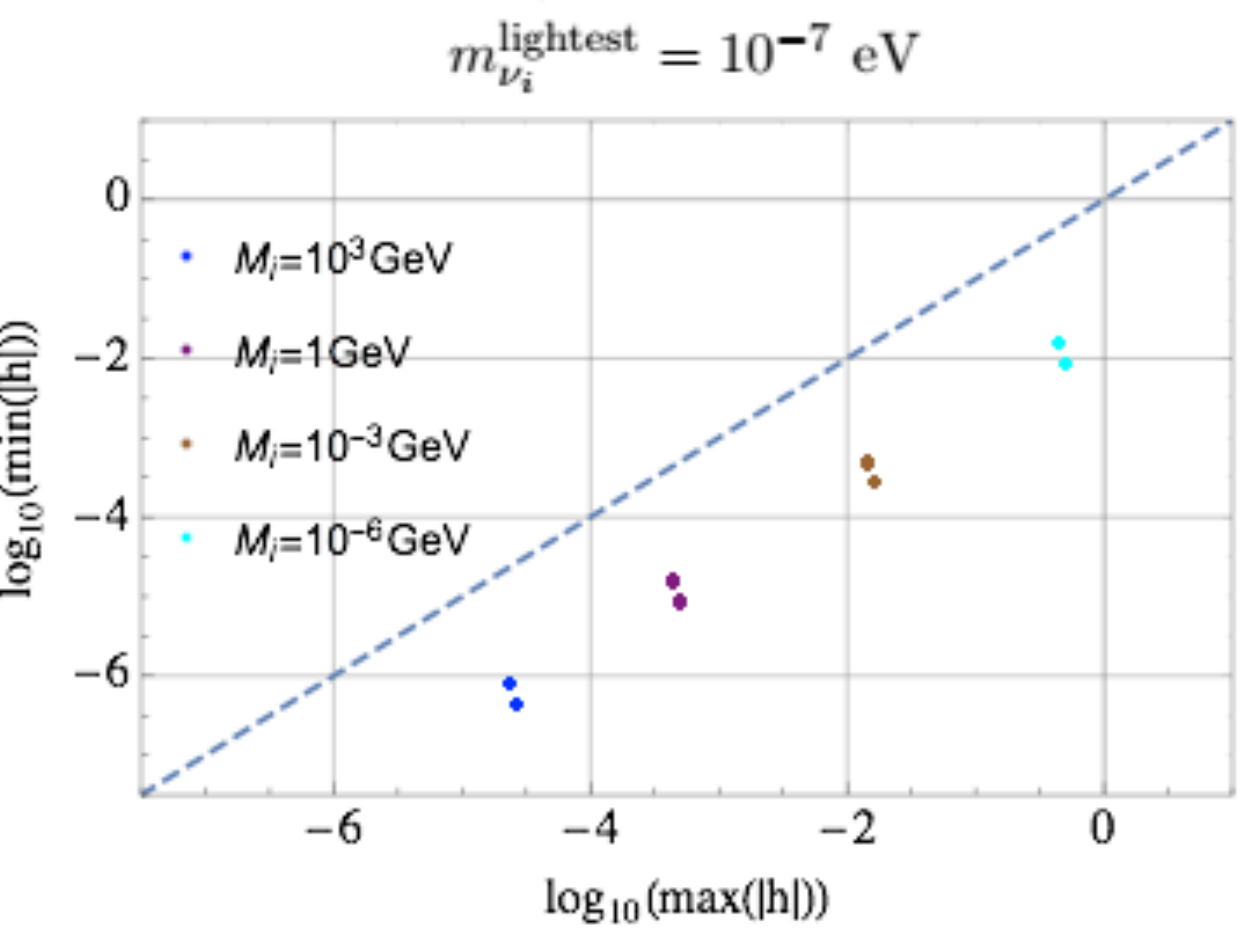}
	\includegraphics[width=0.49\textwidth]{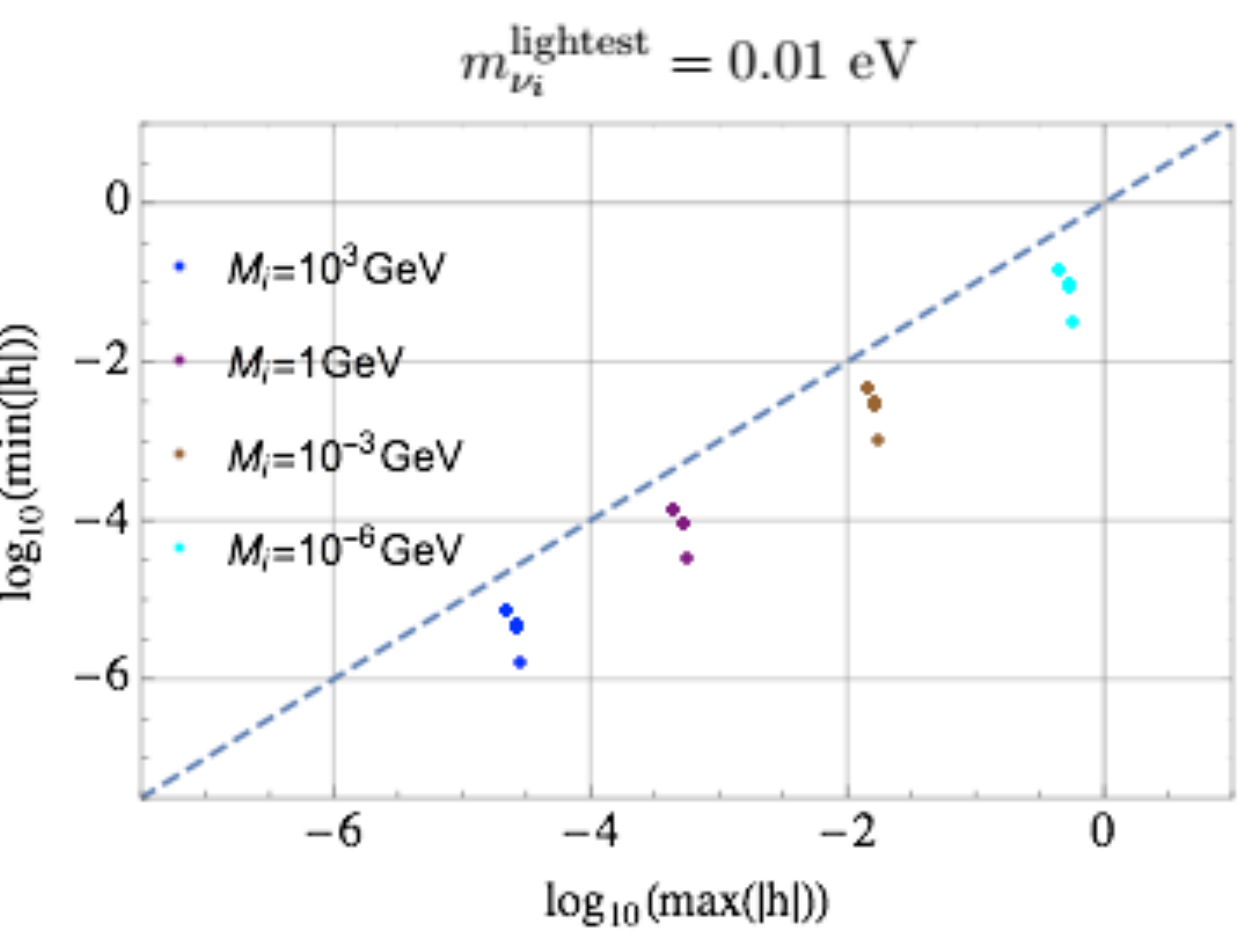}
	\caption{Case $M_i < \Lambda_i$: Allowed points in terms of max$ (|h|) $ versus min$ (|h|) $ for various $ M_i\equiv M_1 = M_2 = M_3 $ values. The loop-induced neutrino masses are from Eq.~\eqref{top neutrino mass}. The dashed line corresponds to maximally anarchic HY matrix, i.e. max$ (|h|) =$ min$ (|h|) $.  In the left and right panel, the mass of the lightest neutrino is $ m_{\nu_i}^{\rm lightest}= 10^{-7} $ and $ 0.01 $ eV, respectively, for both NH and IH. }
	\label{minmaxh 1}
\end{figure}

\begin{figure}[t!]
	\centering
	\includegraphics[width=0.49\textwidth]{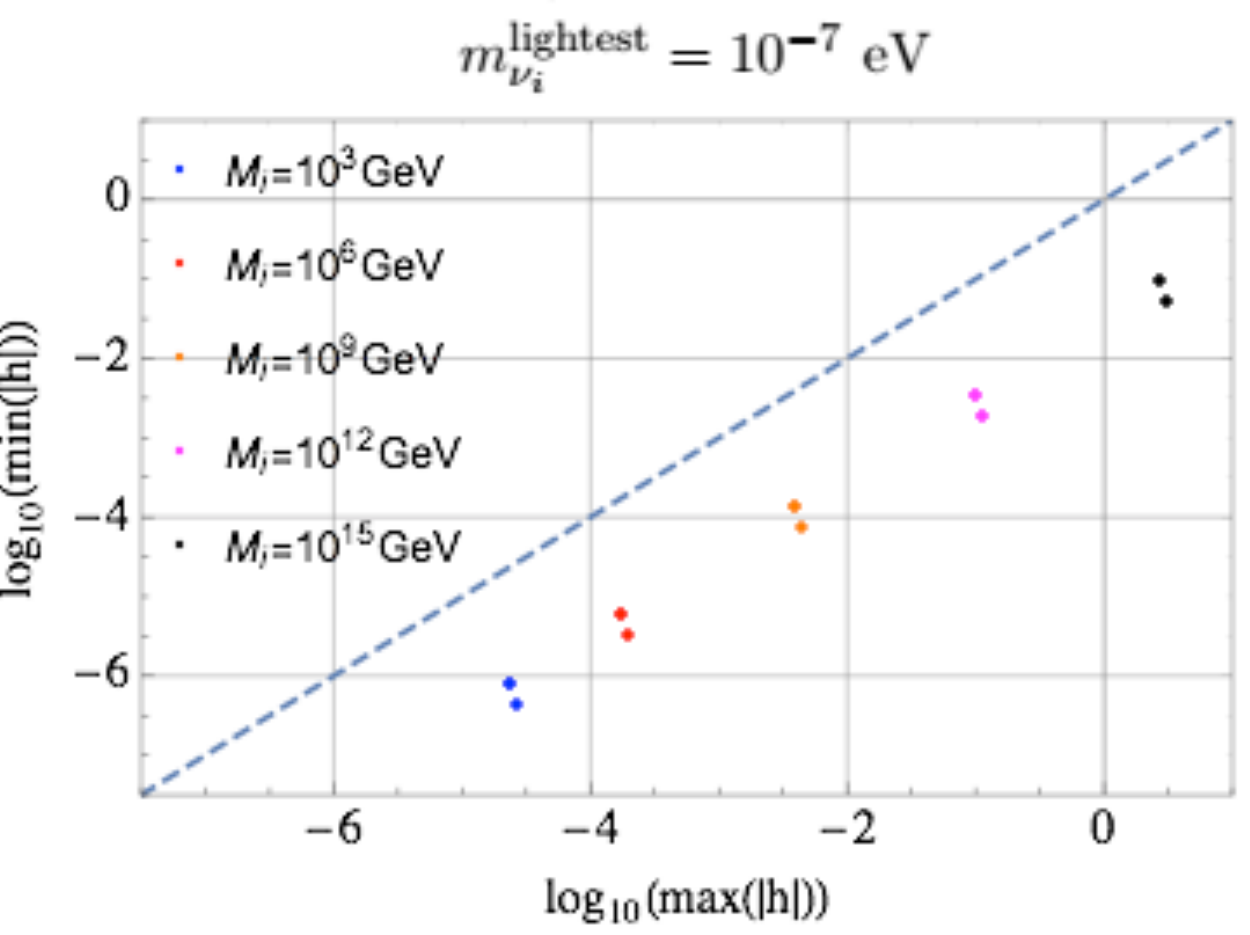}
	\includegraphics[width=0.49\textwidth]{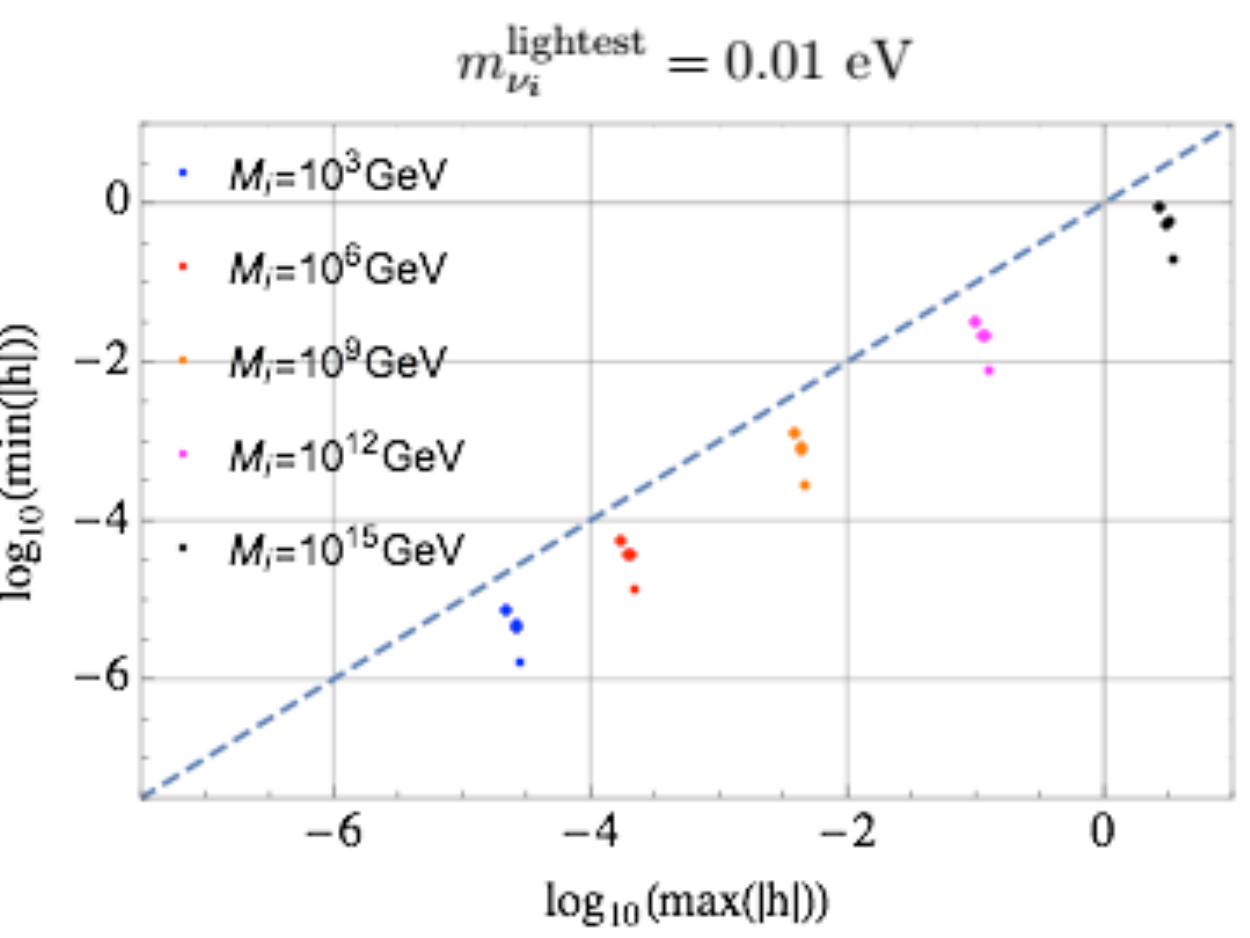}
	\caption{Case $M_i > \Lambda_i$: Same as Fig.~\ref{minmaxh 1}, for the loop-induced neutrino mass in Eq.~\eqref{top neutrino mass Mi large} for ETC and FPC case, and in Eq.~\eqref{top neutrino mass PC} for PC case where $ M_i $ are replaced by $ m_\ast^a $.}
	\label{minmaxh 2}
\end{figure}

Before discussing the values of neutrino masses, we need to explore how degenerate are the two mass eigenstates $ \rm Re\widetilde{\eta}^0 $ and $ \rm Im\widetilde{\eta}^0 $ with the masses $ m_{R/I} $, respectively: for this purpose, we define $ \Delta_{\eta}\equiv (m_R-m_I)/m_I $. We recall that $\Delta_\eta =0$ corresponds to vanishing neutrino masses in Eqs.~(\ref{top neutrino mass})-(\ref{top neutrino mass PC}), so that a small mass split induces a further suppression on top of the loop. We thus open up the parameter space of the model by allowing $C_{LR}$ and $s_\theta$ to vary: in Fig.~\ref{ReImEta0 masses}, we show its dependence on $- C_{\rm LR}$ for two values of $s_\theta$.
The high level of degeneracy in the masses results in reduced neutrino masses from loop effects in Eq.~(\ref{top neutrino mass}) ($ M_i < \Lambda_{\rm i} $), in Eq.~(\ref{top neutrino mass PC}) ($ M_i \gg \Lambda_{\rm i} $ for PC), or in Eq.~(\ref{top neutrino mass Mi large}) ($ M_i \gg \Lambda_{\rm i} $ for ETC and FPC). Above a certain value of $ -C_{\rm LR} $, $ \Delta_{\eta} $ becomes complex due to the instability of the vacuum. 
The mass degeneracy is enhanced close to the boundary of the instability (thus requiring fine tuning to keep it stable), while it remains constant well below. Thus, we consider reasonable to fix $ -C_{\rm LR}=1 $ for simplicity, like in our benchmark.  It should also be noted that reducing $s_\theta$ also increases the degeneracy: we could obtain the correct neutrino masses for order unity $h^{ij}$ and right-handed neutrino masses at the TeV scale with $s_\theta \approx 10^{-5}$. This small value of the misalignment angle, however, requires a strong fine tuning in the potential generated by the top loops. 

In Figs.~\ref{minmaxh 1} and \ref{minmaxh 2}, valid for the neutrino masses in Eqs.~\eqref{top neutrino mass},~(\ref{top neutrino mass Mi large}), and~\eqref{top neutrino mass PC}, respectively, we show allowed values of the neutrino HY couplings in terms of max$ (|h|) $ versus min$ (|h|) $ for various values of the degenerate right-handed neutrino masses, $ M_i\equiv M_1 = M_2 = M_3 $, where $ h\equiv h^{ij} $ or $ h\equiv h^{ai} $ are the neutrino Yukawa coupling constants. However, in the PC case where $ M_i \gg m_\ast^a  $, the results in the figures are still valid except that the right-handed neutrino masses, $ M_i $, are replaced by the masses of their degenerate composite partners, $ m_\ast^a \equiv m_\ast^1=m_\ast^2 = m_\ast^3 $. For both figures, in the left and right panel, the mass of the lightest neutrino is fixed to $ 10^{-7} $ and $ 0.01 $ eV, respectively, for both NH and IH. Firstly, the difference between max$ (|h|) $ and min$ (|h|) $ are decreasing for increasing mass of the lightest neutrino, $ m_{\nu_i}^{\rm lightest} $, due to the fact that the mass-squared differences of the neutrinos are fixed to their measured values (Eq.~(\ref{best-fit experimental values}) for NH or Eq.~(\ref{best-fit experimental values IH}) for IH). Moreover, this behavior appears in the figures by the fact that the colored dots are approaching the dashed line (where max$ (|h|)=\text{min}(|h|) $) for increasing $ m_{\nu_i}^{\rm lightest} $, meaning that the differences between ${\rm max}(\vert h\vert )$ and ${\rm min}(\vert h\vert) $ are reducing. Therefore, we have chosen the largest of these two values in the following, namely $ m_{\nu_1}=0.01 $ eV for NH, because we are interested in the maximal anarchic HY matrix. Secondly, the order of magnitude of the Yukawa coupling constants are increasing for increasing $  M_i $ above $ M_i \sim 10^3 $ GeV or decreasing $  M_i $ below $ M_i \sim 10^3 $ GeV. Finally, by setting $ m_3=0 $ does not change much of the above conclusions.

In the following, we provide some numerical examples for the neutrino Yukawas, for the three approaches considered in Section~\ref{IIIB}. In these examples, we will consider the case without a CP violating Dirac phase, $ \delta =0 $. Generally, if $ \delta \neq 0 $, the following results do not change significantly except that almost all HY coupling constants are complex. For simplicity, we will also consider the NH case only, because the IH case leads to the similar conclusions. \\

\textbf{(i) ETC-type four-fermion operators:} By assuming $ \Lambda_\nu = 50 \text{ TeV}\sim 5 \Lambda_{\rm HC} $, $ M_i = 10^{15} $ GeV, and $ N=2 $ and $ A=1 $ in Eq.~(\ref{condense Yukawa operators}), there exists one positive, real solution of the four-fermion coupling constants $  y^{ij}_\nu$ in Eq.~(\ref{four fermion operators neutrinos}) with maximal number of zeros by using Eq.~(\ref{top neutrino mass Mi large}) for $ M_i \gg \Lambda_\nu $: \beq
y_\nu^{ij}= \begin{pmatrix}1.66 &0.36 &0.34\\0 &2.14  &1.66 \\ 0 & 0 & 2.98 \end{pmatrix}.
\eeq

\begin{figure}[t!]
	\centering
	\includegraphics[width=0.7\textwidth]{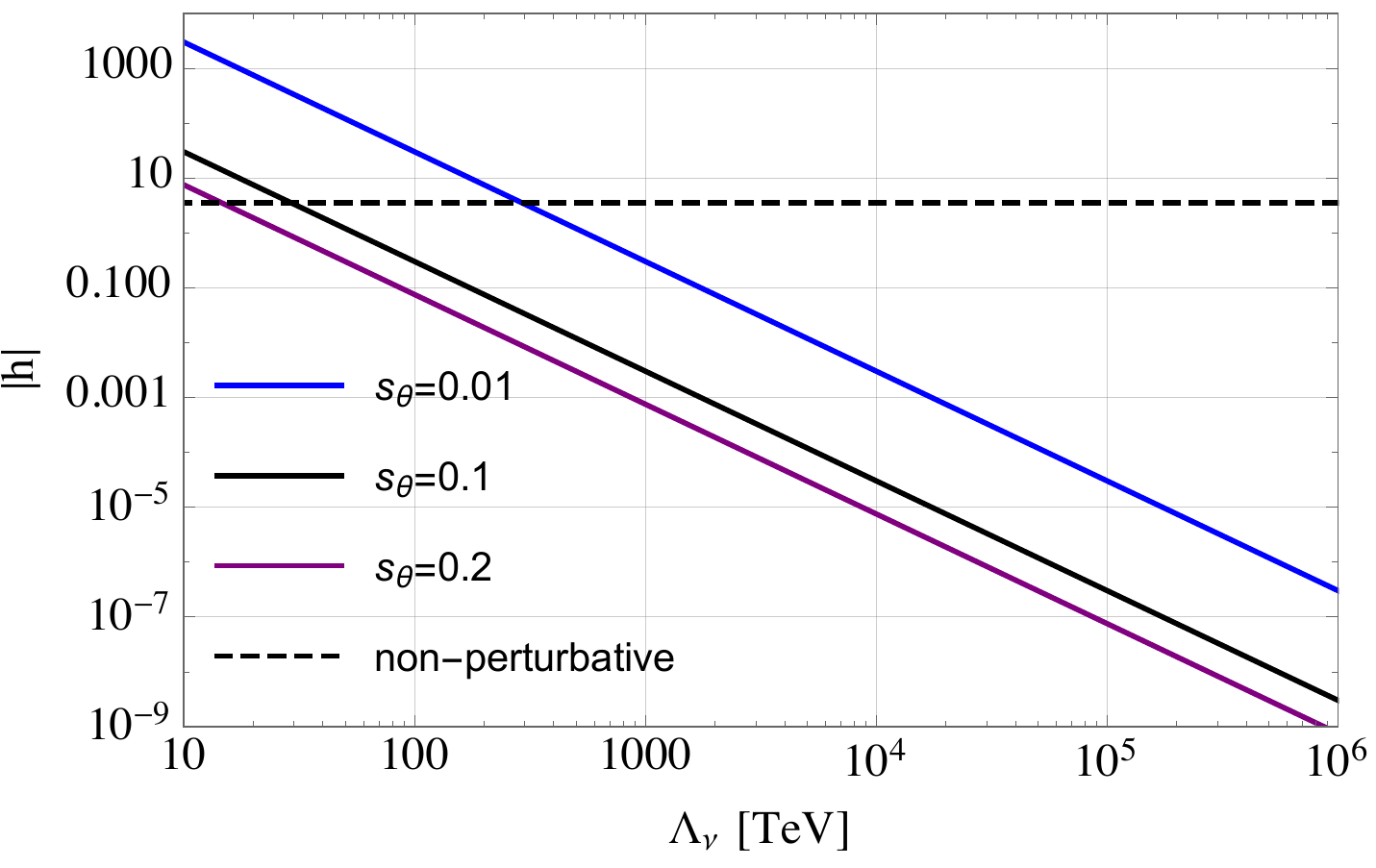}
	\caption{The Yukawa couplings $ \vert h\vert $ for varying UV scale $ \Lambda_\nu $ in Eq.~(\ref{four fermion operators neutrinos}) for $ s_\theta = 0.01,0.10,0.20  $. The dashed line represents the upper limit for perturbative couplings, $ \vert h\vert \lesssim \sqrt{4\pi} $.  }
	\label{hLambdanu}
\end{figure}

Another possibility could be to fix $ \vert y^{ij}_\nu\vert = \mathcal{O}(1) $ by increasing the UV scale $ \Lambda_\nu $ in Eq.~(\ref{four fermion operators neutrinos}) instead of $ M_i $ as shown in Fig.~\ref{hLambdanu}. For example, if we assume that $ M_i = 1000 $ GeV and $ \Lambda_\nu = 1.5\times 10^7 $ GeV, by using Eq.~(\ref{top neutrino mass}) for $ M_i< \Lambda_\nu$, we obtain one positive, real solution of the Yukawa couplings with maximal number of zeros: \beq
y_\nu^{ij}= \begin{pmatrix}1.26 &0.27 &0.26\\0 & 1.62  &1.26 \\ 0 & 0 &2.26 \end{pmatrix}.
\eeq 

\textbf{(ii) Fermion fundamental partial compositeness:} By assuming $ M_S = 50 \text{ TeV}\sim 5 \Lambda_{\rm HC}$, $ M_i=10^{15} $ GeV, and the Yukawa couplings $ y_\nu^{ij}\equiv y_N^{ij} $ in Eq.~(\ref{yukawa couplings ii}), one positive, real solution of the Yukawa couplings exists: \beq y_\nu^{ij}\equiv y_N^{ij} = \begin{pmatrix}1.29 &0.13 &0.09\\0 & 1.46  &0.52 \\ 0 & 0 &1.73 \end{pmatrix},
\eeq  where we have used Eq.~(\ref{top neutrino mass Mi large}) for $ M_i \gg M_S $.

In this approach, we also have another possibility, where we fix $ \vert y_{L,N}^{ij}\vert = \mathcal{O}(1) $ by adjusting $ M_S $ instead of $ M_i $. We can consider a similar example as in the ETC-type approach with $ M_i=1000 $ GeV and $ M_S = 1.5\times 10^7 $ GeV. In this case, there is one positive, real solution of the Yukawa couplings with maximal zeros:\beq y_\nu^{ij}\equiv y_N^{ij} = \begin{pmatrix}1.12 &0.11 &0.08\\0 & 1.27  &0.45 \\ 0 & 0 &1.50 \end{pmatrix},
\eeq where we have used Eq.~(\ref{top neutrino mass}) for $ M_i < M_S $. \\

\textbf{(iii) Fermion partial compositeness:} We assume $ M_i=1000 $ GeV, $ m_\ast^a =\Lambda_{\rm HC}\approx 10.9 $ TeV, and $ \Lambda_{\rm UV}=100 $ TeV in Eq.~(\ref{eff yukawa couplings iii specific}). By using Eq.~(\ref{top neutrino mass}) for $ M_i< m_\ast^a$, we obtain that one of the positive, real solutions of the Yukawa couplings, $ h^{ij} $, leads to the following specific hierarchy of $ \epsilon_i^\psi $ in Eq.~(\ref{eff yukawa couplings iii specific}): \begin{equation}
\begin{aligned}
& \epsilon_1^L=0.31, \quad \epsilon_2^L=1.70, \quad \epsilon_3^L=1.04, \quad \\ & \epsilon_1^N=0.27, \quad \epsilon_2^N=1.48, \quad \epsilon_3^N=0.90,  \quad 
\end{aligned} \end{equation} where the strong-sector low-energy coupling $ g_*=1.00 $ and $ c_{ab}\sim \mathcal{O}(1) $. 


In the case with $ M_i \gg m_*^a $, we obtain positive, real solutions of the Yukawa couplings, $ h^{ai} $, each leading to a specific hierarchy of $ \epsilon_i^\psi $ in Eq.~(\ref{eff yukawa couplings iii specific 2}). For example, if $ M_i = 10^5 $ TeV, $ m_\ast^a=\Lambda_{\rm HC}\approx 10.9 $ TeV, and $ \Lambda_{\rm UV}=m_\ast^a $ (no walking dynamics) by using Eq.~(\ref{top neutrino mass PC}), we obtain that \begin{equation}
\begin{aligned}
& \epsilon_1^L=0.52, \quad \epsilon_2^L=2.80, \quad \epsilon_3^L=1.71, \quad \\ & \epsilon_1^N=1.02, \quad \epsilon_2^N=2.36, \quad \epsilon_3^N=1.85,  \quad 
\end{aligned} \end{equation} where $ g_*=1.00 $ and $ c_{ab}\sim \mathcal{O}(1) $. 

By adding similar terms as in Eq.~(\ref{yukawa couplings iii}) including the charged SM fermions already added for the top quark in Eq.~(\ref{eq: top PC operators}), we can estimate the values of $ \epsilon_a^\psi $ for the charged SM fermions. These values are shown in Table~1 in Ref.~\cite{Frigerio:2018uwx}. All these values of $ \epsilon_a^\psi $ give rise to the observed SM fermion masses and mixing.  \\

In the three approaches, the operators that generate the neutrino masses and mixing also give rise to contributions in the misalignment potential, similar to the terms in Eqs.~(\ref{eq: V top p2}) and~(\ref{eq: V top p4}) from the top PC operators in Eq.~(\ref{eq: top PC operators}). These extra potential contributions induce a dependence of the masses $ m_{R,I} $ on $ h^{ij} $. However, for large $ M_i\gg \Lambda_{\rm HC} $ (or for $ M_i\sim \Lambda_{\rm HC} $), the extra potential contributions from both ETC and FPC-type approach are proportional to $ \Lambda_{\rm HC}/M_i $ (or $ (h^{ij})^2 $), while for PC-type approach the $ y_{N}^2 $ term (similar to the $ y_R^2 $ term in the top PC) is also proportional to $ \Lambda_{\rm HC}/M_i $ (or $ (h^{ij})^2 $) and the $ y_{N}^4 $ terms $ (\Lambda_{\rm HC}/M_i)^2 $ (or $ (h^{ij})^4 $). These contributions to the masses $ m_{R,I} $ are negligible, because either $ M_i $ are very large compared to $ \Lambda_{\rm HC} $ or $ h^{ij} $ are very small. The similar $ y_{\nu}^2 $ term to the $ y_L^2 $ term from the top PC is not allowed if the left-handed neutrinos are in the symmetric representation of the chiral group $ \SU(6) $. Finally, the $ y_{\nu}^4 $ terms for the PC-type approach are suppressed by $ (h^{ij})^4 $, when $  M_i \sim \Lambda_{\rm HC} $, but they are not suppressed by large $ M_i $. However, they have negligible effects on the masses $ m_{R,I} $ even for $ M_i \gg \Lambda_{\rm HC} $. However, we have included all these extra contributions in the above calculations. 

\section{Experimental constraints} \label{exp constraints}

In the following, we discuss the experimental constraints of this model. \\

\begin{figure}[t!]
	\centering
	\includegraphics[width=0.7\textwidth]{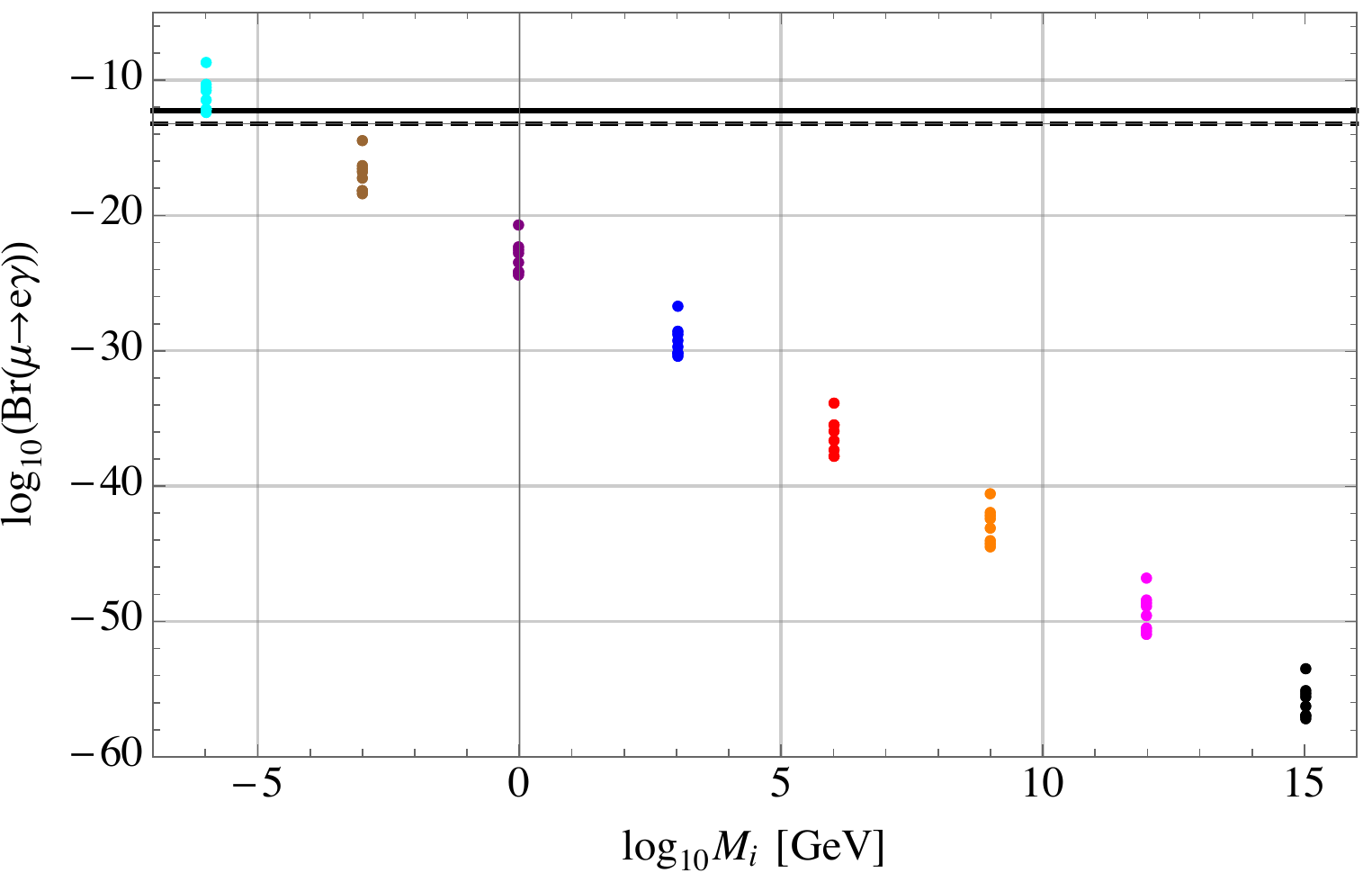}
	\caption{ $ \text{Br}(\mu\rightarrow e\gamma) $ as a function of $ M_i $ for $ m_3=m_2=m_1 $ and $ s_\theta =0.1 $. The solid and dashed line represent the present bound~\cite{Adam:2013mnn} and the future sensitivity~\cite{Baldini:2013ke}, respectively. The various colors of the points represent the same values of $ M_i $ as in Figs.~\ref{minmaxh 1} and~\ref{minmaxh 2}.   }
	\label{LFV processes}
\end{figure}

\textbf{Lepton flavour violating (LFV) processes:} LFV decay processes occur at one-loop level from exchange of $ N_{R,i} $ and $ \widetilde{\eta}^\pm $, where the field $ \widetilde{\eta}^\pm $ is the mass eigenstate of the mass matrix $ M_{\pm}^2 $ in the basis $ (\eta^\pm, \Delta^\pm) $ consisting of mostly $ \eta^\pm $. We will include the experimental constraints from the LFV decays $ l_\alpha \rightarrow l_\beta + \gamma $ and $ l_\alpha \rightarrow l_\beta + \overline{l}_\beta + l_\beta $. The branching ratio for $ l_\alpha \rightarrow l_\beta + \gamma $ is given by~\cite{Toma:2013zsa} \beq \label{eq: LFV BR llgamma}\text{Br}(l_\alpha \rightarrow l_\beta + \gamma)= \frac{3\alpha_{\rm EM} v_{\rm EW}^4}{32 \pi m_{\widetilde{\eta}^\pm}^4}\Bigg \vert \sum_{k=1}^3 h_{\beta k}^* h_{\alpha k} F(M_k^2/m_{\widetilde{\eta}^\pm}^2)\Bigg\vert^2, \nonumber
\eeq where $ \alpha_{\rm EM}=e^2/4\pi $ is the electromagnetic fine structure constant and $ F(x)=(1-6x+3x^2+2x^3-6x^2 \log x)/6(1-x)^4 $. The expression of the branching ratio for $ l_\alpha \rightarrow l_\beta + \overline{l}_\beta + l_\beta $ is given in Ref.~\cite{Toma:2013zsa}. In Fig.~\ref{LFV processes}, we show the branching ratio for the LFV decay process $ \mu \rightarrow e+\gamma $ in our model, as a function of $ M_i $ for $ m_3=m_2=m_1 $ and $ s_\theta = 0.1 $. The solid and dashed line represent the present bound~\cite{Adam:2013mnn} and the future sensitivity~\cite{Baldini:2013ke}, respectively. 
Only very small masses for the right-handed neutrinos are disfavoured, giving a lower bound $ M_i \gtrsim 10^{-6} $ GeV, while  the other LFV decay processes $ l_\alpha \rightarrow l_\beta + \gamma $ and $ l_\alpha \rightarrow l_\beta + \overline{l}_\beta + l_\beta $ give weaker constraints.  \\

\textbf{The electroweak precision tests:} The contributions from the Higgs loops in this model to the electroweak precision parameters (EWPTs) $ S $ and $ T $ are~\cite{Cacciapaglia:2020kgq}  \beq 
&&\Delta S = \frac{1}{6\pi}\left[(1-c_\theta^2)\ln\frac{\Lambda_{\rm HC}}{m_h}+N_D s_\theta^2\right], \\
&&\Delta T = -\frac{3}{8\pi c^2_{\theta_W}}\left[(1-c_\theta^2)\ln \frac{\Lambda_{\rm HC}}{m_h}\right],
\eeq where $ N_D $ is the number of hyper-fermion doublets. The bounds from the EWPTs are $ S=0.06\pm 0.09 $ and $ T=0.10\pm 0.07 $ for $ U=0 $ with correlation $ 0.91 $~\cite{Baak:2014ora}. These bounds only give rise to an upper bound on $ s_\theta $. The upper bound on $ s_\theta $ for $ G_{\rm HC}=\SU(2)_{\rm HC} $ and $ G_{\rm HC}=\SP(4)_{\rm HC} $ with two hyper-fermion doublets as in the concrete $ \SU(6)/\SP(6) $ model are $ s_\theta < 0.24 $ and $ s_\theta < 0.20 $, respectively. Therefore, $ s_\theta=0.1 $ in our above benchmark is viable. \\

\textbf{The ratio $ R_{\gamma\gamma} $ and $ R_{\gamma Z} $:} The particle $ \widetilde{\eta}^{\pm} $ modifies the value of the branching ratio $\text{Br}(h\rightarrow \gamma\gamma)$ at loop level. The combined results from CMS and ATLAS collaborations on the ratio $ R_{\gamma \gamma}\equiv \text{Br}(h\rightarrow \gamma\gamma)/\text{Br}(h\rightarrow \gamma\gamma)_{\rm SM} =1.09\pm 0.12 $~\cite{ATLAS:2017ovn}, while the ratio $ R_{\gamma Z} $ is not measured yet. The expressions for $ R_{\gamma \gamma} $ and $ R_{\gamma Z} $ are given in Ref.~\cite{Chen:2013vi}. This gives rise to no constraint on the vacuum alignment angle $ s_\theta $ for $ m_3=m_2=m_1 $ and a constraint $ s_\theta \lesssim 0.95 $ for $ m_3=0 $ from the measurements of $ R_{\gamma \gamma} $. This is much weaker than the constraint from the EW precision tests. \\

\textbf{Gauge boson decay widths:}  Finally, the following conditions should be fulfilled to keep the W and Z gauge boson decay modes unmodified:  \beq 
m_R+m_I>&& m_Z, \quad\quad\quad m_{\widetilde{\eta}^\pm}+m_R > m_W, \quad \nonumber \\
m_{\widetilde{\eta}^\pm}+m_I >&& m_W, \quad\quad\quad 2 m_{\widetilde{\eta}^\pm}>m_Z.  \nonumber
\eeq These conditions have been successfully met in this model because the masses $ m_{R,I} $ and $ m_{\widetilde{\eta}^\pm} $ are all about 1.1 TeV for $ s_\theta =0.1$; even for the maximum value of $ s_\theta $ from the EW precision tests, they are met. \\

\textbf{Dark matter relic density:} The lightest $ \mathbb{Z}_2 $--odd particle in the particle spectrum can become the DM candidate which in this model is either in the form of the lightest right-handed neutrino or the lightest $ \mathbb{Z}_2 $--odd composite scalar particle. According to the latest data from the Planck satellite in Ref.~\cite{Ade:2015xua}, DM accounts for around 26\% of our Universe mass budget at present. The density parameter of DM is~\cite{Ade:2015xua} \beq \label{exp DM abundance}
\Omega_{\rm DM}h^2 = 0.1186\pm 0.0020. 
\eeq

Initially, we consider the lightest right-handed neutrino as the DM candidate, assuming $ N_{R,1} $, which may be lighter than the lightest $ \mathbb{Z}_2 $--odd composite scalar, namely the mass eigenstate $ \widetilde{\varphi}^0 $ of the mass matrix $ M_{I}^2 $ in the basis $ (\rm Im\eta^0, \varphi^0) $ consisting mostly of $ \varphi^0 $. In this case, the mass of $ N_{R,1} $ may be lighter than $ m_{\widetilde{\varphi}^0}=661 $ GeV for $ s_\theta=0.1 $. 

Firstly, we consider this candidate as cold dark matter (CDM). The relic density of CDM depends on the thermally averaged cross section $ \langle \sigma_{\rm eff}\vert v_{\rm rel}\vert\rangle $. In this model, the thermally averaged cross section is computed from annihilations of the lightest right-handed neutrino into the left-handed neutrinos and charged leptons via t-channel diagrams mediated by the members of the composite $ \mathbb{Z}_2 $--odd doublet, and can be written as $ \langle \sigma_{\rm eff}\vert v_{\rm rel}\vert\rangle = a_{\rm eff} + 6 b_{\rm eff}/x$, where~\cite{Kong:2005hn} \beq 
 a_{\rm eff}=&&\frac{1}{16\pi}\frac{M_1^2}{(M_1^2+m_0^2)^2}\sum_{ij} (h^{i1}h^{j2}-h^{i2}h^{j1})^2, \nonumber \\ 
b_{\rm eff}=&&\frac{1}{16\pi}\frac{M_1^2}{(M_1^2+m_0^2)^2}\frac{m_0^4-3m_0^2M_1^2-M_1^4}{3(M_1^2+m_0^2)^2}  \sum_{ij}(h^{i1}h^{j2}-h^{i2}h^{j1})^2 + \\ 
&&\frac{1}{48\pi}\frac{M_1^2(M_1^4+m_0^4)}{(M_1^2+m_0^2)^4} \sum_{ij}(h^{i1}h^{j1}+h^{i2}h^{j2})^2  \nonumber
\eeq with $ m_0 \equiv (m_R+m_I)/2 $. The relic abundance of CDM can be estimated by~\cite{Kong:2005hn} \beq 
\Omega_{N_1}h^2 = \frac{107\times 10^9 x_f}{g_*^{1/2}m_{P}(\text{GeV}) (a_{\rm eff}+3b_{\rm eff})/x_f},
\eeq where $ m_P=1.22\times 10^{19} $ GeV, $ g_*=106.75 $, the freeze-out parameter is \beq 
x_f\equiv \frac{M_1}{T_f} = \ln \left[\frac{0.038 g_{\rm eff}m_P M_1 \langle \sigma_{\rm eff}\vert v_{\rm rel}\vert\rangle}{g_*^{1/2}x_f^{1/2}}\right],
\eeq and \beq 
g_{\rm eff}=\sum_{i=1}^{3} g_{N_{R,i}}(1+\Delta_i)^{3/2}e^{-\Delta_i x} \nonumber
\eeq with $ \Delta_i =(M_i-M_1)/M_1 $ depicting the mass splitting ratio of $ N_{R,i} $, and $ g_{N_{R,i}} $ are the number of degrees of freedom of $ N_{R,i} $. If one of the right-handed neutrinos is CDM candidate, then it is overproduced for all possible masses compared to the DM abundance given by Eq.~(\ref{exp DM abundance}). It is the smallness of the Yukawa couplings $ h^{ij} $ in this mass range of $ M_i $ shown in Figs.~\ref{minmaxh 1} and \ref{minmaxh 2} that results in a small annihilation cross section, and therefore too much relic density of DM. As mention in Section~\ref{Num results}, we can obtain $ h^{ij}=\mathcal{O}(1) $ by tuning the vacuum misalignment angle down to $ s_{\theta}\sim 10^{-5} $, but this leads to heavier composite particles with approximate masses: $ m_0 \sim 100/s_\theta $ GeV ($ m_0 \sim 10^7 $ GeV for $ s_{\theta}\sim 10^{-5} $). This leads again to a small annihilation cross section and an overproduction of DM even for $ M_1 \sim m_0 $. Finally, in the case with $ M_i \gg m_\ast^a $ for the PC approach, we will also have an overproduction of DM by assuming the lightest of the composite neutrino partners as DM candidate due to the smallness of $ h^{ai} $. 

Secondly, we consider the lightest $ N_{R,i} $ is light enough to be hot DM. The relic abundance of such species can be calculated simply by following the standard prescription given by E.~W.~Kolb and M.~S.~Turner in Ref.~\cite{Kolb:1990vq}, the present abundance of $ N_{R,1} $ as hot DM can be written as \beq 
\Omega_{N_{R,1}}h^2 = 57.3 \frac{g_{N_{R,1}}}{g_{*S}(x_f)}\frac{M_1}{\rm keV},
\eeq where $ g_{N_{R,1}} $ is the number of degrees of freedom of $ N_{R,1} $, and $ g_{*S}(x_f) $ represents the number of relativistic entropy degrees of freedom at the epoch of $ N_{R,1} $ decoupling at $ x_f $. If $ N_{R,1} $ should play the role as hot DM, its mass should be \beq 
0.01 \text{ keV} < M_1 < 0.11  \text{ keV},
\eeq because the value of $ g_{*S}(x_f) $ is in the range from 10.75 if $ N_{R,1} $ decouples after the QCD phase transition to 107 if the decoupling occurs above the EWSB scale. If the mass $  M_1 $ is larger than the upper limit, there will be an overproduction of DM. This kind of DM is not viable either because they do not meet the constraints by the LFV processes in Fig.~\ref{LFV processes} which requires $ M_i \gtrsim 1 $ keV for the maximum value of $ s_\theta $.  For smaller $ s_\theta $, the constraints of the LFV processes are even stronger, and thus the lower limit of $ M_i $ is larger. Therefore, the masses of $ N_{R,i} $ may be heavier than the lightest $ \mathbb{Z}_2 $--odd composite scalar to avoid overproduction of DM. 

Thus, the last possibility for a DM candidate in this model can be the lightest $ \mathbb{Z}_2 $--odd composite scalar. This possibility has been investigated in Refs.~\cite{Cai:2018tet,Cai:2019cow}. In Ref.~\cite{Cai:2019cow}, it has been shown that the lightest of the $\mathbb{Z}_2$--odd composite scalars may provide the correct DM relic density via non-thermal asymmetric production.  \\

\textbf{The XENON1T excess:} In our paper~\cite{Cacciapaglia:2020kbf}, we identified the parameter space in a composite scenario where the light pseudo-scalar $ \eta $, identical to the $ \chi $ state in Table~\ref{tab:su6sp6}, can be produced in the sun and explain the XENON1T excess in electron recoil data~\cite{Aprile:2020tmw}. We considered the $ \SU(6)/\SP(6) $ template model as here, where the misalignment angle may be in the range $ 0.004 < s_\theta < 0.007 $ to explain this excess. The model's DM candidate, arising in a non-thermal way, has a mass around $ 50 $ TeV and out of range for Direct Detection. Even for such small $ s_\theta $, this model can still explain the smallness of the neutrino masses with the mechanism explained in this paper. Finally, according to Ref.~\cite{Cacciapaglia:2020kbf}, this model with such small $ s_\theta $ gives rise to additional testable predictions include gravitation waves at frequencies in the Hz range from a cosmological phase transition, an exotic decay $Z \to \gamma + \mbox{inv.}$ with rates $4 \div 16 \cdot 10^{-12}$ testable at a future Tera-Z collider, and an enhancement by $17\div 40$\% of the branching ratio $K_L \to \pi^0 + \mbox{inv.}$, not enough to explain the KOTO anomaly~\cite{Ahn:2018mvc}. All these predictions may be confirmed by future experiments. 

\section{Conclusions}

We have presented a novel mechanism to generate small neutrino masses in composite Higgs models. The mechanism, similar in nature to the scotogenic models, naturally features two suppression mechanisms: neutrino masses are loop generated via the coupling to $\mathbb{Z}_2$--odd composite pNGBs; the near-degeneracy of the pNGBs results in a further suppression. Thus, even for sizeable couplings to the composite sector, neutrinos can obtain small enough masses.  This mechanism can also be featured in a wide variety of other models based on vacuum misalignment. 

We have considered an $ \SU(6)/\SP(6) $ CH template, which naturally features two composite Higgs doublets, one of which can be made inert. In this template model, we have investigated three different approaches to generate the neutrino Yukawa couplings: (i) ETC-type four-fermion operators, (ii) ``Fundamental Partial Compositeness'', and (iii) ``Partial Compositeness''. These three approaches can give rise to Yukawa coupling constants of order unity for masses of the right-handed neutrinos in the mass range from the lightest $\mathbb{Z}_2$--odd composite particle ($ \sim $TeV scale for $ s_\theta=0.1 $) up to the Planck scale (however only $ \sim 10^{18} $ GeV for $ s_\theta=0.1 $). Therefore, these scale limits depend on the vacuum misalignment angle $\theta $. The lower limit originates from the Dark Matter relic density: we have demonstrated that the right-handed neutrinos will result in overproduction in the Universe if they are lighter than the $\mathbb{Z}_2$--odd composite scalars. 
The upper bound originates from the upper limit for the perturbative couplings, $ \vert h \vert \lesssim \sqrt{4\pi} $. 
Finally, we have checked various experimental constraints for these three approaches, showing that no strong constraints arise (except for the DM one). When the lightest of the $\mathbb{Z}_2$--odd state is a composite scalar, it can provide the correct relic density either by the usual thermal freeze-out or as a non-thermal asymmetric relic, as shown in Ref.~\cite{Cai:2019cow}. 

The composite scotogenic mechanism, therefore, can provide a natural explanation of the lightness of neutrinos in various models based on vacuum misalignment for the EW symmetry breaking, while also featuring a composite (a)symmetric DM candidate presented in Ref.~\cite{Cai:2019cow} and, as presented in Ref.~\cite{Cacciapaglia:2020kbf}, an explaination of the recent XENON1T excess in electron recoil data~\cite{Aprile:2020tmw}.

\section*{Acknowledgements}
G.C. acknowledges partial support from the Labex-LIO (Lyon Institute of Origins) under grant ANR-10-LABX-66 (Agence Nationale pour la Recherche), and FRAMA (FR3127, F\'ed\'eration de Recherche ``Andr\'e Marie Amp\`ere''). 
MR acknowledges partial funding from The Council For Independent Research, grant number DFF 6108-00623. The CP3-Origins centre is partially funded by the Danish National Research Foundation, grant number DNRF90.

\appendix 

\section{Neutrino mass matrices for $ M_k\gg \Lambda_i $} \label{sec: Appendix A}

Here, we will derive the expressions of the neutrino mass matrix for $ M_k\gg \Lambda_i $ for the three approaches with (i) ETC-type four-fermion operators, (ii) ``Fundamental Partial Compositeness'', and (iii) ``Partial Compositeness''. \\

\begin{figure}[t!]
	\centering
	\includegraphics[width=1.0\textwidth]{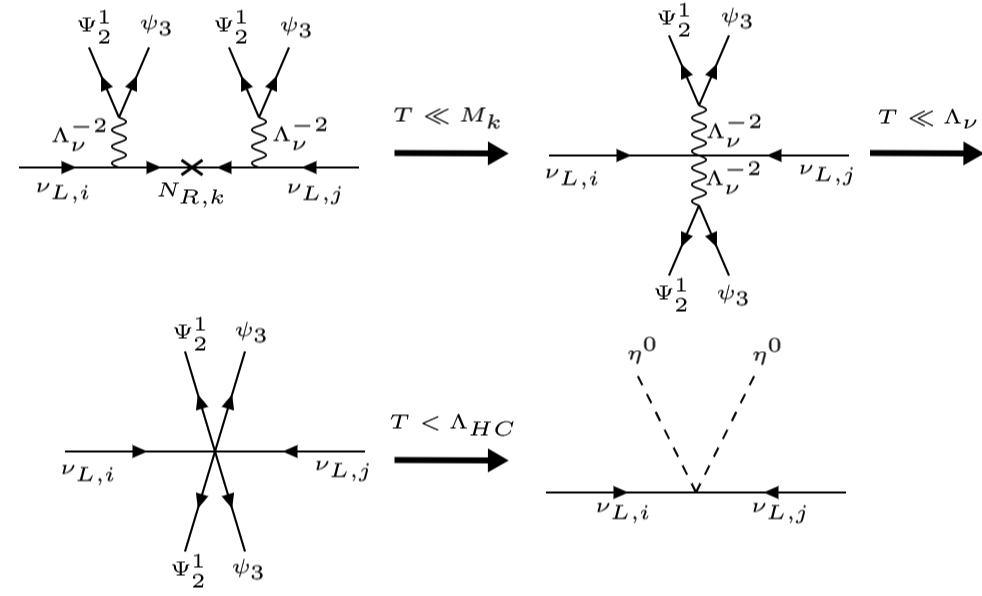}
	\caption{ Development of the $ (\eta^0)^2 (\nu_{L})^2 $ four-vertices for the ETC-type approach when $ M_i \gg \Lambda_\nu $ from temperatures above $ M_k $ down to zero-temperature by integrating out the propagators of the right-handed neutrinos, $ N_{R,k} $, and the ETC gauge or scalar boson(s) with masses in order of $ \Lambda_\nu $. In this process, $ M_k \gg \Lambda_\nu \gg \Lambda_{\rm HC} $. }
	\label{fig: Developing of four-vertex ETC}
\end{figure}

\textbf{(i) ETC-type four-fermion operators:} In the case where $ M_i\gg \Lambda_\nu $, the ETC-type four-fermion operators for the neutrinos in Eq.~(\ref{four fermion operators neutrinos}) are replaced by six-fermion operators (valid for the temperatures $ \Lambda_{\rm HC} < T\ll \Lambda_\nu  $)\begin{equation}
\begin{aligned}
&\sum_{k=1}^3\frac{y_\nu^{ik}y_\nu^{kj}}{\Lambda_\nu^4 M_k}(\Psi^1_2 \psi_3)(\Psi^1_2 \psi_3)\nu_{L,i}^\dagger \nu_{L,j}^\dagger,
\end{aligned}
\end{equation} which are developed after the first two steps in Fig.~\ref{fig: Developing of four-vertex ETC}. In this process, the propagators of the right-handed neutrinos, $ N_{R,k} $, with masses $ M_k $ and the ETC gauge or scalar boson(s) with masses in order of $ \Lambda_\nu $ are integrated out. 
When the hyper-fermions condense below $ \Lambda_{\rm HC} $, these six-fermion operators are replaced by the four-vertices after the last step in Fig.~\ref{fig: Developing of four-vertex ETC}:
\begin{equation}
\begin{aligned} \label{eq: Weinberg operator ETC}
&\sum_{k=1}^3 \frac{y_\nu^{ik}y_\nu^{kj}}{\Lambda_\nu^4 M_k}(\Psi^1_2 \psi_3)(\Psi^1_2 \psi_3)\nu_{L,i}^\dagger \nu_{L,j}^\dagger  \rightarrow  \sum_{k=1}^3 \frac{h^{ik}h^{kj}}{M_k}(\eta^0)^2 \nu_{L,i}^\dagger \nu_{L,j}^\dagger
\end{aligned}
\end{equation} with $ h^{ij}\equiv 4\pi N A (\Lambda_{\rm HC}/\Lambda_\nu)^2 y_\nu^{ij}$~\cite{Hill:2002ap}, where $ A $ is an integration constant arising from the condensation and $ N $ is the number of hyper-colors. The definition of $ h^{ij} $ here is identical to the definition in Eq.~(\ref{eq: Yukawa coupling ETC FPC PC}) for $ M_i < \Lambda_\nu $.

Therefore, for $ M_i\gg \Lambda_\nu $, the one-loop diagrams in Fig.~\ref{loopdiagram2} contribute to the SM neutrino masses instead of the diagrams in Fig.~\ref{loopdiagram} for $ M_k < \Lambda_\nu $. The radiative mass contribution with a $\rm Re\widetilde{\eta}^0 $ loop is \begin{equation}
\begin{aligned} \label{eq: loop calculation appendix}
-i\Sigma_{ijk}^R=&\int \frac{d^4p}{(2\pi)^4}i\frac{h^{ik}h^{kj}}{M_k}\frac{i}{p^2-m_R^2+i\epsilon}\\ 
=&- i \frac{h^{ik}h^{kj}}{(4\pi)^2 M_k}\left[\Lambda_{\rm HC}\sqrt{\Lambda_{\rm HC}^2+m_R^2}+m_R^2 \ln\left(\frac{m_R}{\Lambda_{\rm HC}+\sqrt{\Lambda_{\rm HC}^2+m_R^2}}\right)\right], 
\end{aligned}
\end{equation} where we have used $ \Lambda_{\rm HC} $ as cut-off scale due to the fact that the $ (\eta^0)^2 (\nu_{L})^2 $ four-vertices will dissolve above this scale. There is also another diagram with an $\rm Im\widetilde{\eta}^0 $ loop which comes with opposite sign. Therefore, in the ETC case for $ M_k \gg \Lambda_\nu $, the neutrino mass matrix as in Eq.~(\ref{top neutrino mass Mi large}) can be expressed as \begin{equation}
\begin{aligned} \label{eq:neutrino mass matrix}
m_{\nu,ij}=&\sum_{k=1}^3\left(\Sigma_{ijk}^R+\Sigma_{ijk}^I \right)\\ =&\sum_{k=1}^{3} \frac{h^{ik}h^{kj}}{(4\pi)^2 M_k}\Bigg[\Lambda_{\text{HC}}\sqrt{m_R^2+\Lambda_{\text{HC}}^2}+ m_R^2 \ln\Bigg(\frac{m_R}{\Lambda_{\text{HC}}+\sqrt{m_R^2+\Lambda_{\text{HC}}^2}}\Bigg) \\&\phantom{\sum_{k=1}^{3} \frac{h^{ki}h^{kj}}{(4\pi)^2 M_k}\Bigg[} -\Lambda_{\text{HC}}\sqrt{m_I^2+\Lambda_{\text{HC}}^2}- m_I^2 \ln\Bigg(\frac{m_I}{\Lambda_{\text{HC}}+\sqrt{m_I^2+\Lambda_{\text{HC}}^2}}\Bigg) \Bigg]. 
\end{aligned}
\end{equation} 

\begin{figure}[t!]
	\centering
	\includegraphics[width=1.0\textwidth]{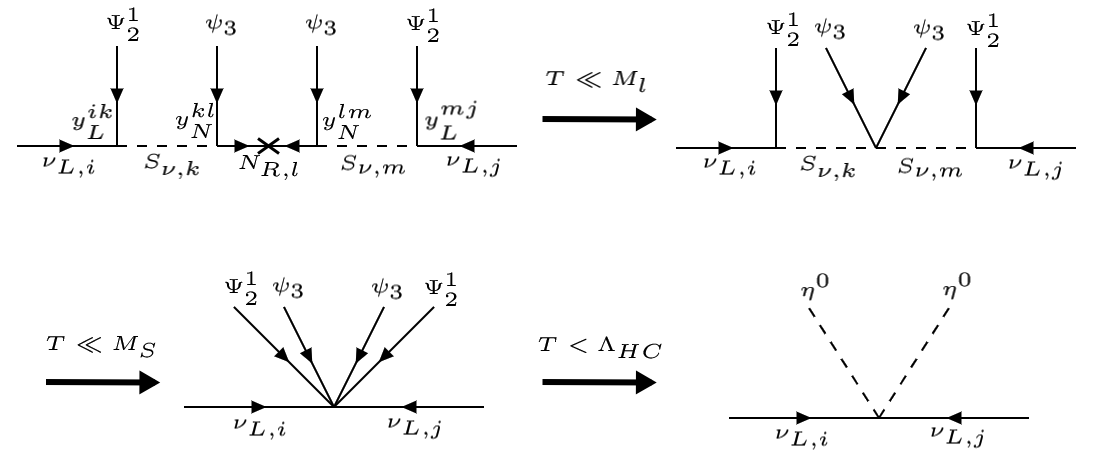}
	\caption{Development of the $ (\eta^0)^2 (\nu_{L})^2 $ four-vertices for the FPC approach when $ M_i \gg M_S $ from temperatures above $ M_i $ down to zero-temperature by integrating out the propagators of the right-handed neutrinos, $ N_{R,l} $, with masses $ M_l $ and the techni-scalars, $ S_{\nu,k} $ and $ S_{\nu,m} $, with the mass $ M_S $. In this process, $ M_i \gg M_S \gg \Lambda_{\rm HC} $. }
	\label{fig: Developing of four-vertex FPC}
\end{figure}

\textbf{(ii) Fermion fundamental partial compositeness:} In the case where $ M_i\gg M_S $, the FPC fundamental Lagrangian terms for the neutrinos in Eq.~(\ref{yukawa couplings ii}) are replaced by six-fermion operators (valid for the temperatures $ \Lambda_{\rm HC}<T\ll M_S $) \begin{equation}
\begin{aligned}
&\sum_{k,l,m=1}^3\frac{y_L^{ik}y_N^{kl}y_N^{lm}y_L^{mj}}{M_S^4 M_l}(\Psi^1_2 \psi_3)(\Psi^1_2 \psi_3)\nu_{L,i}^\dagger \nu_{L,j}^\dagger,
\end{aligned}
\end{equation} which are developed after the first two steps in Fig.~\ref{fig: Developing of four-vertex FPC}. In this process, the propagators of the right-handed neutrinos, $ N_{R,l} $, with masses $ M_l $ and the techni-scalars, $ S_{\nu,k} $ and $ S_{\nu,m} $, with the mass $ M_S $ are integrated out. 
When the hyper-fermions condense below $ \Lambda_{\rm HC} $, these six-fermion operators generate the four-vertices after the last step in Fig.~\ref{fig: Developing of four-vertex FPC}:
\begin{equation}
\begin{aligned} \label{eq: Weinberg operator FPC}
&\sum_{k,l,m=1}^3\frac{y_L^{ik}y_N^{kl}y_N^{lm}y_L^{mj}}{M_S^4 M_l}(\Psi^1_2 \psi_3)(\Psi^1_2 \psi_3)\nu_{L,i}^\dagger \nu_{L,j}^\dagger  \rightarrow \sum_{k=1}^3\frac{h^{ik}h^{kj}}{M_k}(\eta^0)^2 \nu_{L,i}^\dagger \nu_{L,j}^\dagger, 
\end{aligned}
\end{equation} with $ h^{ij}\equiv 4\pi N C_{\text{Yuk}} (\Lambda_{\rm HC}/M_S)^2 y_L^{ik}y_N^{kj}$~\cite{Hill:2002ap}, where $ C_{\text{Yuk}} $ is an integration constant arising from the condensation and $ N $ is the number of hyper-colors. The definition of $ h^{ij} $ here is identical to the definition in Eq.~(\ref{eq: Yukawa coupling ETC FPC PC}) for $ M_i< M_S $. Therefore, the form of the $ (\eta^0)^2 (\nu_{L})^2 $ four-vertices are identical for both the ETC-type and FPC approach for large $ M_i\gg \Lambda_\nu,M_S $. Therefore, for $ M_i \gg M_S $, the neutrino mass matrix has the same form as in Eq.~(\ref{eq:neutrino mass matrix}), while the form of $  h^{ij} $ in Eq.~(\ref{eq: Weinberg operator FPC}) differs from the form of $ h^{ij} $ in Eq.~(\ref{eq: Weinberg operator ETC}) for ETC-type operators. \\

\textbf{(iii) Fermion partial compositeness:} 
We recall from Eq.~(\ref{PCeffeciveops}) that the PC four-fermion interactions of the neutrinos generated at $\Lambda_{\rm UV}$ are given by \beq  \label{four-fermion operators PC appendix}
\frac{y_\nu^{ia}(\Lambda_{\rm UV})}{\Lambda_{\rm UV}^2} \nu^{\dagger}_{L,i} (\Psi^\dagger P_L^1 \Psi^*\chi_{\nu,a}^\dagger) + \frac{y_N^{ia}(\Lambda_{\rm UV})}{\Lambda_{\rm UV}^2}N_{R,i}^\dagger (\Psi^\dagger P_N \Psi \chi_{\nu,a}^\dagger)+\rm h.c. \ \ \ \ \ \
\eeq 

\begin{figure}[t!]
	\centering
	\includegraphics[width=0.8\textwidth]{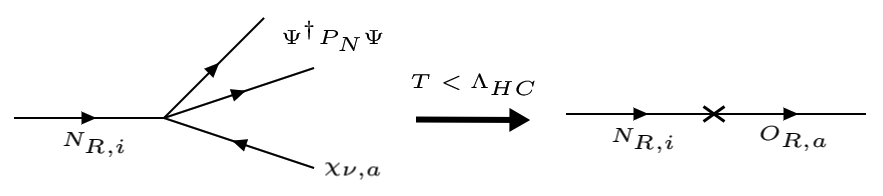}
	\caption{ The generation of the mixing terms of the right-handed neutrinos, $ N_{R,i} $, and their composite partners, $ O_{R,a} $, from the four-fermion interactions in Eq.~(\ref{four-fermion operators PC appendix}) when $ M_i< m_* $.}
	\label{fig: Majorana Masses small Appendix}
\end{figure}

\begin{figure}[t!]
	\centering
	\includegraphics[width=1.0\textwidth]{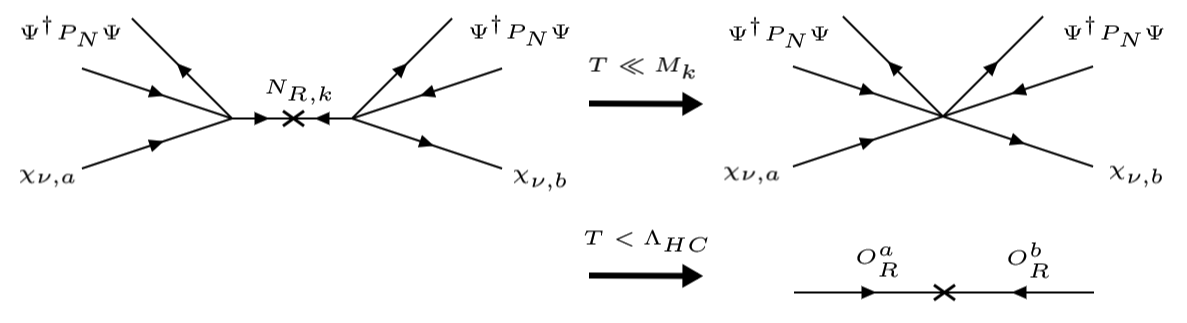}
	\caption{The generation of the Majorana masses of the right-handed composite neutrino partners, $ O_{R,a} $, from the four-fermion interactions in Eq.~(\ref{four-fermion operators PC appendix}) when $ M_i\gg m_* $.}
	\label{fig: Majorana Masses large Appendix}
\end{figure} 

Firstly, upon condensation for $ M_i < m_*^a $, we obtain (illustrated in Fig.~\ref{fig: Majorana Masses small Appendix}) \beq  
 \frac{y_N^{ia}(\Lambda_{\rm UV})}{\Lambda_{\rm UV}^2} (m_*^a)^3 N_{R,i}^\dagger O_{R,a}\simeq   \epsilon^N_{ia} m_*^a N_{R,i}^\dagger O_{R,a},
\eeq  where $ O_{L,R}^a $ are the left- and right-handed composite neutrino partners, and the couplings $ y_{\nu,N} $ are scaled from $ \Lambda_{\rm UV} $ down to $ m_*\equiv m_*^a $ as follows \beq 
y_{\nu,N}(m_*)\simeq y_{\nu,N}(\Lambda_{\rm UV})\left(\frac{m_*}{\Lambda_{\rm UV}}\right)^2\equiv \epsilon_{\nu,N}.  
\eeq 

Secondly, upon condensation for $ M_i \gg m_*^a $, the propagators of the right-handed neutrinos will be integrated out before condensation as illustrated in the first step in Fig.~\ref{fig: Majorana Masses large Appendix}. Thus, the right-handed composite neutrino partners ($ O_{R,a} $) achieve Majorana mass terms,  \begin{equation}
\begin{aligned}
\sum_{k=1}^3 \frac{(m_*^a)^3(m_*^b)^3}{M_k \Lambda_{\rm UV}^4} y_N^{ak}(\Lambda_{\rm UV})y_N^{kb}(\Lambda_{\rm UV}) O_{R,a}O_{R,b}
& \simeq \sum_{k=1}^3  m_*^a\frac{\epsilon_{ak}^N \epsilon_{kb}^N}{M_k} m_*^b O_{R,a}O_{R,b}.
\end{aligned}
 \end{equation}
 
\begin{figure}[t!]
	\centering
	\includegraphics[width=1.0\textwidth]{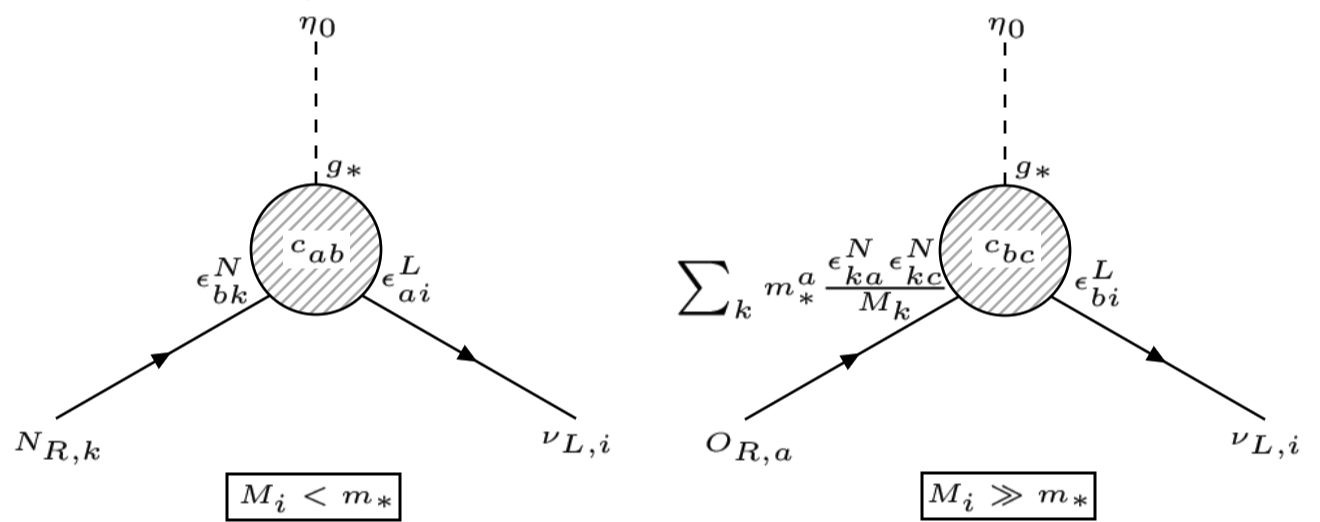}
	\caption{Diagrams for the PC approach showing how Yukawa couplings with the neutrinos and $ \eta^0 $ are formed for $ M_i<m_* $ and $ M_i\gg m_* $, respectively.  }
	\label{fig: Yukawa PC Appendix}
\end{figure}

 
Thus, the neutrino Yukawa couplings for $ M_i < m_*^a $ and $ M_i \gg m_*^a $ are given by\begin{align}
 \label{eff yukawa couplings iii specific light appendix}
&h_{ik}\equiv \frac{A_N-\sqrt{2}B_N}{16\sqrt{2}\pi}  g_* \sum_{a,b=1}^3 (\epsilon_{ia}^L)^* c_{ab} \epsilon_{bk}^N \quad \quad \text{for} \ \ M_i < m_*^a, \\ \label{eff yukawa couplings iii specific heavy appendix}
&h_{ai}\equiv \frac{A_N-\sqrt{2}B_N}{16\sqrt{2}\pi}  g_* \sum_{k,b,c=1}^3 \frac{ m_\ast^a}{M_k}\epsilon_{ak}^N \epsilon_{kc}^N c_{cb} (\epsilon_{bi}^L)^*  \quad\quad \text{for} \ \ M_i \gg m_*^a,
 \end{align} which are formed in the way as illustrated in Fig.~\ref{fig: Yukawa PC Appendix}. These couplings are, respectively, the neutrino Yukawa couplings in Eq.~(\ref{eff yukawa couplings iii specific}) for $ M_i < m_*^a $ and the modified couplings of Eq.~(\ref{eff yukawa couplings iii specific 2}) for $ M_i \gg m_*^a $, where the right-handed neutrinos, $ N_{R,i} $, are replaced by the right-handed composite neutrino partners, $ O_{R}^a $.

The one-loop diagrams for the neutrino masses, similar to the diagrams in Fig.~\ref{loopdiagram} where $ N_{R,k} $ are replaced with $ O_{R,a} $, give rise to the mass expression \begin{equation}
\begin{aligned} \label{top neutrino mass PC appendix}
m_{\nu}^{ij}=&&\sum_{a=1}^3 \frac{h^{ia}h^{aj}}{(4\pi)^2}m_\ast^a\Bigg[\frac{m_R^2}{m_R^2-(m_\ast^a)^2}\ln\left(\frac{m_R^2}{(m_\ast^a)^2}\right) -\frac{m_I^2}{m_I^2-(m_\ast^a)^2}\ln\left(\frac{m_I^2}{(m_\ast^a)^2}\right)\Bigg],
\end{aligned}
 \end{equation} where the modified couplings $ h^{ai} $ are given in Eq.~(\ref{eff yukawa couplings iii specific heavy appendix}).


\begin{thebibliography}{99}

\bibitem{Weinberg:1967tq}
S.~Weinberg,
``A Model of Leptons,''
\href{http://dx.doi.org/10.1103/PhysRevLett.19.1264}{Phys. Rev. Lett. \textbf{19} (1967), 1264-1266}.

\bibitem{Pontecorvo:1967fh}
B.~Pontecorvo,
``Neutrino Experiments and the Problem of Conservation of Leptonic Charge,''
Sov. Phys. JETP \textbf{26} (1968), 984-988.

\bibitem{Minkowski:1977sc}
P.~Minkowski,
``$\mu \to e\gamma$ at a Rate of One Out of $10^{9}$ Muon Decays?,''
\href{http://dx.doi.org/10.1016/0370-2693(77)90435-X}{Phys. Lett. B \textbf{67} (1977), 421-428}.

\bibitem{Mohapatra:1979ia}
R.~N.~Mohapatra and G.~Senjanovic,
``Neutrino Mass and Spontaneous Parity Nonconservation,''
\href{http://dx.doi.org/10.1103/PhysRevLett.44.912}{Phys. Rev. Lett. \textbf{44} (1980), 912}.

\bibitem{Yanagida:1980xy}
T.~Yanagida,
``Horizontal Symmetry and Masses of Neutrinos,''
\href{http://dx.doi.org/10.1143/PTP.64.1103}{Prog. Theor. Phys. \textbf{64} (1980), 1103}.

\bibitem{Ma:2006km}
E.~Ma,
``Verifiable radiative seesaw mechanism of neutrino mass and dark matter,'' Phys. Rev. D \textbf{73} (2006), [\href{https://arxiv.org/abs/hep-ph/0601225}{{\tt hep-ph/0601225}}].

\bibitem{Appelquist:2002me}
T.~Appelquist and R.~Shrock,
``Neutrino masses in theories with dynamical electroweak symmetry breaking,''
Phys. Lett. B \textbf{548} (2002), 204-214, [\href{https://arxiv.org/abs/hep-ph/0204141}{{\tt hep-ph/0204141}}].

\bibitem{Appelquist:2003hn}
T.~Appelquist, M.~Piai and R.~Shrock,
``Fermion masses and mixing in extended technicolor models,''
Phys. Rev. D \textbf{69} (2004), 015002, [\href{https://arxiv.org/abs/hep-ph/0308061}{{\tt hep-ph/0308061}}].

\bibitem{Bellazzini:2014yua}
B.~Bellazzini, C.~Cs\'{a}ki and J.~Serra,
``Composite Higgses,''
\href{http://dx.doi.org/10.1140/epjc/s10052-014-2766-x}{Eur. Phys. J. C \textbf{74} (2014) no.5, 2766}, [\href{https://arxiv.org/abs/1401.2457}{{\tt 1401.2457}}].

\bibitem{Panico:2015jxa}
G.~Panico and A.~Wulzer,
``The Composite Nambu-Goldstone Higgs,''
\href{http://dx.doi.org/10.1007/978-3-319-22617-0}{Lect. Notes Phys. \textbf{913} (2016), pp.1-316}, [\href{https://arxiv.org/abs/1506.01961}{{\tt 1506.01961}}].

\bibitem{Cacciapaglia:2020kgq}
G.~Cacciapaglia, C.~Pica and F.~Sannino,
``Fundamental Composite Dynamics: A Review,''
\href{http://dx.doi.org/0.1016/j.physrep.2020.07.002}{Phys. Rept. \textbf{877} (2020), 1-70}, [\href{https://arxiv.org/abs/2002.04914}{{\tt 2002.04914}}].

\bibitem{Holdom:1981rm}
B.~Holdom,
``Raising the Sideways Scale,''
\href{http://dx.doi.org/10.1103/PhysRevD.24.1441}{Phys. Rev. D \textbf{24} (1981), 1441}.

\bibitem{Carmona:2013cq}
A.~Carmona and F.~Goertz,
``Custodial Leptons and Higgs Decays,''
\href{http://dx.doi.org/10.1007/JHEP04(2013)163}{JHEP \textbf{04} (2013), 163}, [\href{https://arxiv.org/abs/1301.5856}{{\tt 1301.5856}}].

\bibitem{Carmona:2014iwa}
A.~Carmona and F.~Goertz,
``A naturally light Higgs without light Top Partners,''
\href{http://dx.doi.org/10.1007/JHEP05(2015)002}{JHEP \textbf{05} (2015), 002}, [\href{https://arxiv.org/abs/1410.8555}{{\tt 1410.8555}}].

\bibitem{Carmona:2015ena}
A.~Carmona and F.~Goertz,
``Lepton Flavor and Nonuniversality from Minimal Composite Higgs Setups,''
\href{http://dx.doi.org/10.1103/PhysRevLett.116.251801}{Phys. Rev. Lett. \textbf{116} (2016) no.25, 251801}, [\href{https://arxiv.org/abs/1510.07658}{{\tt 1510.07658}}].

\bibitem{Frigerio:2018uwx}
M.~Frigerio, M.~Nardecchia, J.~Serra and L.~Vecchi,
``The Bearable Compositeness of Leptons,''
\href{http://dx.doi.org/10.1007/JHEP10(2018)017}{JHEP \textbf{10} (2018), 017}, [\href{https://arxiv.org/abs/1807.04279}{{\tt 1807.04279}}].

\bibitem{Kaplan:1983fs}
D.~B. Kaplan and H.~Georgi, ``{SU(2) x U(1) Breaking by Vacuum
  Misalignment},''\href{http://dx.doi.org/10.1016/0370-2693(84)91177-8}{\emph{Phys.
  Lett.} {\bf 136B} (1984) 183--186}.

\bibitem{Galloway:2016fuo}
J.~Galloway, A.~L. Kagan and A.~Martin, ``{A UV complete partially
  composite-pNGB
  Higgs},''\href{http://dx.doi.org/10.1103/PhysRevD.95.035038}{\emph{Phys.
  Rev.} {\bf D95} (2017) 035038}, [\href{https://arxiv.org/abs/1609.05883}{{\tt
  1609.05883}}].

\bibitem{Alanne:2017rrs}
T.~Alanne, D.~B.~Franzosi and M.~T. Frandsen, ``{A partially composite
  Goldstone
  Higgs},''\href{http://dx.doi.org/10.1103/PhysRevD.96.095012}{\emph{Phys.
  Rev.} {\bf D96} (2017) 095012}, [\href{https://arxiv.org/abs/1709.10473}{{\tt
  1709.10473}}].

\bibitem{Alanne:2017ymh}
T.~Alanne, D.~Buarque~Franzosi, M.~T. Frandsen, M.~L.~A. Kristensen, A.~Meroni
  and M.~Rosenlyst, ``{Partially composite Higgs models: Phenomenology and RG
  analysis},''\href{http://dx.doi.org/10.1007/JHEP01(2018)051}{\emph{JHEP} {\bf
  01} (2018) 051}, [\href{https://arxiv.org/abs/1711.10410}{{\tt 1711.10410}}].


\bibitem{Barducci:2018yer}
D.~Barducci, S.~De Curtis, M.~Redi and A.~Tesi,
``An almost elementary Higgs: Theory and Practice,''
\href{http://dx.doi.org/10.1007/JHEP08(2018)017}{JHEP \textbf{08} (2018), 017}, [\href{https://arxiv.org/abs/1805.12578}{{\tt
  1805.12578}}].

\bibitem{ArkaniHamed:2001nc}
N.~Arkani-Hamed, A.~G. Cohen and H.~Georgi, ``{Electroweak symmetry breaking
  from dimensional
  deconstruction},''\href{http://dx.doi.org/10.1016/S0370-2693(01)00741-9}{\emph{Phys.
  Lett.} {\bf B513} (2001) 232--240},
  [\href{https://arxiv.org/abs/hep-ph/0105239}{{\tt hep-ph/0105239}}].

\bibitem{ArkaniHamed:2002qx}
N.~Arkani-Hamed, A.~G. Cohen, E.~Katz, A.~E. Nelson, T.~Gregoire and J.~G.
  Wacker, ``{The Minimal moose for a little
  Higgs},''\href{http://dx.doi.org/10.1088/1126-6708/2002/08/021}{\emph{JHEP}
  {\bf 08} (2002) 021}, [\href{https://arxiv.org/abs/hep-ph/0206020}{{\tt
  hep-ph/0206020}}].

\bibitem{Contino:2003ve}
R.~Contino, Y.~Nomura and A.~Pomarol, ``{Higgs as a holographic pseudoGoldstone
  boson},''\href{http://dx.doi.org/10.1016/j.nuclphysb.2003.08.027}{\emph{Nucl.
  Phys.} {\bf B671} (2003) 148--174},
  [\href{https://arxiv.org/abs/hep-ph/0306259}{{\tt hep-ph/0306259}}].

\bibitem{Hosotani:2005nz}
Y.~Hosotani and M.~Mabe, ``{Higgs boson mass and electroweak-gravity hierarchy
  from dynamical gauge-Higgs unification in the warped
  spacetime},''\href{http://dx.doi.org/10.1016/j.physletb.2005.04.039}{\emph{Phys.
  Lett.} {\bf B615} (2005) 257--265},
  [\href{https://arxiv.org/abs/hep-ph/0503020}{{\tt hep-ph/0503020}}].

\bibitem{Chacko:2005pe}
Z.~Chacko, H.-S. Goh and R.~Harnik, ``{The Twin Higgs: Natural electroweak
  breaking from mirror
  symmetry},''\href{http://dx.doi.org/10.1103/PhysRevLett.96.231802}{\emph{Phys.
  Rev. Lett.} {\bf 96} (2006) 231802},
  [\href{https://arxiv.org/abs/hep-ph/0506256}{{\tt hep-ph/0506256}}].

\bibitem{Batra:2008jy}
P.~Batra and Z.~Chacko,
``A Composite Twin Higgs Model,''
\href{http://dx.doi.org/10.1103/PhysRevD.79.095012}{Phys. Rev. D \textbf{79} (2009), 095012}, [\href{https://arxiv.org/abs/0811.0394}{{\tt 0811.0394}}].

\bibitem{Barbieri:2015lqa}
R.~Barbieri, D.~Greco, R.~Rattazzi and A.~Wulzer,
``The Composite Twin Higgs scenario,''
\href{http://dx.doi.org/10.1007/JHEP08(2015)161}{JHEP \textbf{08} (2015), 161}, [\href{https://arxiv.org/abs/1501.07803}{{\tt 1501.07803}}].

\bibitem{Low:2015nqa}
M.~Low, A.~Tesi and L.~T.~Wang,
``Twin Higgs mechanism and a composite Higgs boson,''
\href{http://dx.doi.org/10.1103/PhysRevD.91.095012}{Phys. Rev. D \textbf{91} (2015), 095012}, [\href{https://arxiv.org/abs/1501.07890}{{\tt
  1501.07890}}].

\bibitem{Alanne:2014kea}
T.~Alanne, H.~Gertov, F.~Sannino and K.~Tuominen, ``{Elementary Goldstone Higgs
  boson and dark
  matter},''\href{http://dx.doi.org/10.1103/PhysRevD.91.095021}{\emph{Phys.
  Rev.} {\bf D91} (2015) 095021}, [\href{https://arxiv.org/abs/1411.6132}{{\tt
  1411.6132}}].
  
\bibitem{Gertov:2015xma}
H.~Gertov, A.~Meroni, E.~Molinaro and F.~Sannino,
``Theory and phenomenology of the elementary Goldstone Higgs boson,''
\href{http://dx.doi.org/10.1103/PhysRevD.92.095003}{Phys. Rev. D \textbf{92} (2015) no.9, 095003}, [\href{https://arxiv.org/abs/1507.06666}{{\tt
  1507.06666}}].
  
\bibitem{Cai:2018tet}
C.~Cai, G.~Cacciapaglia and H.-H. Zhang, ``{Vacuum alignment in a composite
  2HDM},'' [\href{https://arxiv.org/abs/1805.07619}{{\tt
  1805.07619}}].
  
\bibitem{Cai:2019cow}
C.~Cai, H.-H. Zhang, G.~Cacciapaglia, M.~T. Frandsen and M.~Rosenlyst, ``{Higgs
  emerging from the dark},'' \href{https://arxiv.org/abs/1911.12130}{{\tt
  [1911.12130]}}.

\bibitem{Mrazek:2011iu}
J.~Mrazek, A.~Pomarol, R.~Rattazzi, M.~Redi, J.~Serra and A.~Wulzer,
``The Other Natural Two Higgs Doublet Model,''
\href{http://dx.doi.org/10.1016/j.nuclphysb.2011.07.008}{Nucl. Phys. B \textbf{853} (2011), 1-48}, [\href{https://arxiv.org/abs/1105.5403}{{\tt
  1105.5403}}].

\bibitem{Bertuzzo:2012ya}
E.~Bertuzzo, T.~S.~Ray, H.~de Sandes and C.~A.~Savoy,
``On Composite Two Higgs Doublet Models,''
\href{http://dx.doi.org/10.1007/JHEP05(2013)153}{JHEP \textbf{05} (2013), 153}, [\href{https://arxiv.org/abs/1206.2623}{{\tt
  1206.2623}}].

\bibitem{Ma:2015gra}
T.~Ma and G.~Cacciapaglia,
``Fundamental Composite 2HDM: SU(N) with 4 flavours,''
\href{http://dx.doi.org/10.1007/JHEP03(2016)211}{JHEP \textbf{03} (2016), 211}, [\href{https://arxiv.org/abs/1508.07014}{{\tt
  1508.07014}}].

\bibitem{Ma:2017vzm}
Y.~Wu, T.~Ma, B.~Zhang and G.~Cacciapaglia,
``Composite Dark Matter and Higgs,''
\href{http://dx.doi.org/10.1007/JHEP11(2017)058}{JHEP \textbf{11} (2017), 058}, [\href{https://arxiv.org/abs/1703.06903}{{\tt
  1703.06903}}].

\bibitem{Cacciapaglia:2019ixa}
G.~Cacciapaglia, H.~Cai, A.~Deandrea and A.~Kushwaha,
``Composite Higgs and Dark Matter Model in SU(6)/SO(6),''
\href{http://dx.doi.org/10.1007/JHEP10(2019)035}{JHEP \textbf{10} (2019), 035}, [\href{https://arxiv.org/abs/1904.09301}{{\tt
  1904.09301}}].

\bibitem{Cai:2020njb}
H.~Cai and G.~Cacciapaglia,
``A Singlet Dark Matter in the SU(6)/SO(6) Composite Higgs Model,''
[\href{https://arxiv.org/abs/2007.04338}{\tt 2007.04338}].

\bibitem{Witten:1983tx}
E.~Witten, ``{Current Algebra, Baryons, and Quark
  Confinement},''\href{http://dx.doi.org/10.1016/0550-3213(83)90064-0}{\emph{Nucl.
  Phys.} {\bf B223} (1983) 433--444}.

\bibitem{Kosower:1984aw}
D.~A. Kosower, ``{Symmetry breaking patterns in pseudoreal and real gauge
  theories},''\href{http://dx.doi.org/10.1016/0370-2693(84)91806-9}{\emph{Phys.
  Lett.} {\bf 144B} (1984) 215--216}.

\bibitem{Galloway:2010bp}
J.~Galloway, J.~A. Evans, M.~A. Luty and R.~A. Tacchi, ``{Minimal Conformal
  Technicolor and Precision Electroweak
  Tests},''\href{http://dx.doi.org/10.1007/JHEP10(2010)086}{\emph{JHEP} {\bf
  10} (2010) 086}, [\href{https://arxiv.org/abs/1001.1361}{{\tt 1001.1361}}].

\bibitem{Dugan:1984hq}
M.~J. Dugan, H.~Georgi and D.~B. Kaplan, ``{Anatomy of a Composite Higgs
  Model},''\href{http://dx.doi.org/10.1016/0550-3213(85)90221-4}{\emph{Nucl.
  Phys.} {\bf B254} (1985) 299--326}.


\bibitem{Contino:2010rs}
R.~Contino, ``{The Higgs as a Composite Nambu-Goldstone Boson},'' in
  \emph{{Physics of the large and the small, TASI 09, proceedings of the
  Theoretical Advanced Study Institute in Elementary Particle Physics, Boulder,
  Colorado, USA, 1-26 June 2009}}, pp.~235--306, 2011.
\newblock [\href{https://arxiv.org/abs/1005.4269}{{\tt 1005.4269}}].

\bibitem{Agashe:2006at}
K.~Agashe, R.~Contino, L.~Da~Rold and A.~Pomarol, ``{A Custodial symmetry for
  $Zb \bar
  b$},''\href{http://dx.doi.org/10.1016/j.physletb.2006.08.005}{\emph{Phys.
  Lett.} {\bf B641} (2006) 62--66},
  [\href{https://arxiv.org/abs/hep-ph/0605341}{{\tt hep-ph/0605341}}].

\bibitem{Grojean:2013qca}
C.~Grojean, O.~Matsedonskyi and G.~Panico, ``{Light top partners and precision
  physics},''\href{http://dx.doi.org/10.1007/JHEP10(2013)160}{\emph{JHEP} {\bf
  10} (2013) 160}, [\href{https://arxiv.org/abs/1306.4655}{{\tt 1306.4655}}].

\bibitem{Ghosh:2015wiz}
D.~Ghosh, M.~Salvarezza and F.~Senia, ``{Extending the Analysis of Electroweak
  Precision Constraints in Composite Higgs
  Models},''\href{http://dx.doi.org/10.1016/j.nuclphysb.2016.11.013}{\emph{Nucl.
  Phys.} {\bf B914} (2017) 346--387},
  [\href{https://arxiv.org/abs/1511.08235}{{\tt 1511.08235}}].

\bibitem{BuarqueFranzosi:2018eaj}
D.~Buarque~Franzosi, G.~Cacciapaglia and A.~Deandrea, ``{Sigma-assisted natural
  composite Higgs},'' [\href{https://arxiv.org/abs/1809.09146}{{\tt
  1809.09146}}].

\bibitem{deBlas:2018tjm}
J.~de~Blas, O.~Eberhardt and C.~Krause, ``{Current and Future Constraints on
  Higgs Couplings in the Nonlinear Effective
  Theory},''\href{http://dx.doi.org/10.1007/JHEP07(2018)048}{\emph{JHEP} {\bf
  07} (2018) 048}, [\href{https://arxiv.org/abs/1803.00939}{{\tt 1803.00939}}].

\bibitem{Alanne:2018wtp}
T.~Alanne, N.~Bizot, G.~Cacciapaglia and F.~Sannino,
``Classification of NLO operators for composite Higgs models,''
\href{http://dx.doi.org/10.1103/PhysRevD.97.075028}{Phys. Rev. D \textbf{97} (2018) no.7, 075028}, [\href{https://arxiv.org/abs/1801.05444}{{\tt 1801.05444}}].

\bibitem{Arthur:2016dir}
  R.~Arthur, V.~Drach, M.~Hansen, A.~Hietanen, C.~Pica and F.~Sannino,
  ``SU(2) gauge theory with two fundamental flavors: A minimal template for model building,''
  Phys.\ Rev.\ D {\bf 94} (2016) no.9,  094507, [\href{https://arxiv.org/abs/1602.06559}{{\tt 1602.06559}}].

\bibitem{Dimopoulos:1979es}
S.~Dimopoulos and L.~Susskind,
``Mass Without Scalars,'' \href{https://doi.org/10.1016/0550-3213(79)90364-X}{\emph{Nucl.
  Phys. B} {\bf 155} (1979) 237--252}.  

  \bibitem{Kaplan:1991dc}
D.~B.~Kaplan,
``Flavor at SSC energies: A New mechanism for dynamically generated fermion masses,''
Nucl. Phys. B \textbf{365} (1991), 259-278.

\bibitem{Barnard:2013zea}
J.~Barnard, T.~Gherghetta and T.~S.~Ray,
``UV descriptions of composite Higgs models without elementary scalars,''
\href{http://dx.doi.org/10.1007/JHEP02(2014)002}{JHEP \textbf{02} (2014), 002}, [\href{https://arxiv.org/abs/1311.6562}{{\tt
  1311.6562}}].

\bibitem{Ferretti:2013kya}
G.~Ferretti and D.~Karateev,
``Fermionic UV completions of Composite Higgs models,''
\href{http://dx.doi.org/10.1007/JHEP03(2014)077}{JHEP \textbf{03} (2014), 077}, [\href{https://arxiv.org/abs/1312.5330}{{\tt
  1312.5330}}].

\bibitem{Contino:2011np}
R.~Contino, D.~Marzocca, D.~Pappadopulo and R.~Rattazzi,
``On the effect of resonances in composite Higgs phenomenology,''
JHEP \textbf{10}, 081 (2011), [\href{https://arxiv.org/abs/1109.1570}{{\tt
  1109.1570}}].

\bibitem{Contino:2015mha}
R.~Contino and M.~Salvarezza,
``One-loop effects from spin-1 resonances in Composite Higgs models,''
JHEP \textbf{07}, 065 (2015), [\href{https://arxiv.org/abs/1504.02750}{{\tt
  1504.02750}}].

\bibitem{Matsedonskyi:2012ym}
O.~Matsedonskyi, G.~Panico and A.~Wulzer,
``Light Top Partners for a Light Composite Higgs,''
JHEP \textbf{01}, 164 (2013), [\href{https://arxiv.org/abs/1204.6333}{{\tt
  1204.6333}}].

\bibitem{Redi:2012ha}
M.~Redi and A.~Tesi,
``Implications of a Light Higgs in Composite Models,''
JHEP \textbf{10}, 166 (2012), [\href{https://arxiv.org/abs/1205.0232}{{\tt
  1205.0232}}].

\bibitem{Golterman:2017vdj}
M.~Golterman and Y.~Shamir,
``Effective potential in ultraviolet completions for composite Higgs models,''
Phys. Rev. D \textbf{97}, no.9, 095005 (2018), [\href{https://arxiv.org/abs/1707.06033}{{\tt
  1707.06033}}].

\bibitem{Bennett:2017kga}
Bennett, D.~K.~Hong, J.~W.~Lee, C.~J.~D.~Lin, B.~Lucini, M.~Piai and D.~Vadacchino,
``Sp(4) gauge theory on the lattice: towards SU(4)/Sp(4) composite Higgs (and beyond),''
JHEP \textbf{03}, 185 (2018), [\href{https://arxiv.org/abs/1712.04220}{{\tt
  1712.04220}}].

\bibitem{Lee:2019pwp}
J.~W.~Lee, D.~K.~Hong, C.~J.~D.~Lin, B.~Lucini, M.~Piai, D.~Vadacchino and Bennett,
``Meson spectrum of Sp(4) lattice gauge theory with two fundamental Dirac fermions,''
PoS \textbf{LATTICE2019}, 054 (2019), [\href{https://arxiv.org/abs/1911.00437}{{\tt
  1911.00437}}].


\bibitem{Bennett:2019cxd}
Bennett, D.~K.~Hong, J.~W.~Lee, C.~J.~D.~Lin, B.~Lucini, M.~Mesiti, M.~Piai, J.~Rantaharju and D.~Vadacchino,
``$Sp(4)$ gauge theories on the lattice: quenched fundamental and antisymmetric fermions,''
Phys. Rev. D \textbf{101}, no.7, 074516 (2020), [\href{https://arxiv.org/abs/1912.06505}{{\tt
  1912.06505}}].
  
\bibitem{Ayyar:2017qdf}
V.~Ayyar, T.~DeGrand, M.~Golterman, D.~C.~Hackett, W.~I.~Jay, E.~T.~Neil, Y.~Shamir and B.~Svetitsky,
``Spectroscopy of SU(4) composite Higgs theory with two distinct fermion representations,''
Phys. Rev. D \textbf{97}, no.7, 074505 (2018), [\href{https://arxiv.org/abs/1710.00806}{{\tt
  1710.00806}}].

\bibitem{Ayyar:2018zuk}
V.~Ayyar, T.~Degrand, D.~C.~Hackett, W.~I.~Jay, E.~T.~Neil, Y.~Shamir and B.~Svetitsky,
``Baryon spectrum of SU(4) composite Higgs theory with two distinct fermion representations,''
Phys. Rev. D \textbf{97}, no.11, 114505 (2018), [\href{https://arxiv.org/abs/1801.05809}{{\tt
  1801.05809}}].

\bibitem{Ayyar:2019exp}
V.~Ayyar, M.~F.~Golterman, D.~C.~Hackett, W.~Jay, E.~T.~Neil, Y.~Shamir and B.~Svetitsky,
``Radiative Contribution to the Composite-Higgs Potential in a Two-Representation Lattice Model,''
Phys. Rev. D \textbf{99}, no.9, 094504 (2019), [\href{https://arxiv.org/abs/1903.02535}{{\tt
  1903.02535}}].

\bibitem{Svetitsky:2019hij}
B.~Svetitsky, V.~Ayyar, T.~DeGrand, M.~Golterman, D.~C.~Hackett, W.~I.~Jay, E.~T.~Neil and Y.~Shamir,
``Towards a Composite Higgs and a Partially Composite Top Quark,''
PoS \textbf{LATTICE2019}, 241 (2019), [\href{https://arxiv.org/abs/1911.10867}{{\tt
  1911.10867}}].

\bibitem{Erdmenger:2020flu}
J.~Erdmenger, N.~Evans, W.~Porod and K.~S.~Rigatos,
``Gauge/gravity dual dynamics for the strongly coupled sector of composite Higgs models,''
JHEP \textbf{02}, 058 (2021), [\href{https://arxiv.org/abs/2010.10279}{{\tt
  2010.10279}}].

\bibitem{Kaplan:1983sm}
D.~B.~Kaplan, H.~Georgi and S.~Dimopoulos,
``Composite Higgs Scalars,''
Phys. Lett. B \textbf{136} (1984), 187-190.

\bibitem{Cacciapaglia:2014uja}
G.~Cacciapaglia and F.~Sannino, ``{Fundamental Composite (Goldstone) Higgs
  Dynamics},''\href{http://dx.doi.org/10.1007/JHEP04(2014)111}{\emph{JHEP} {\bf
  04} (2014) 111}, [\href{https://arxiv.org/abs/1402.0233}{{\tt 1402.0233}}].

\bibitem{Hill:2002ap}
C.~T. Hill and E.~H. Simmons, ``{Strong Dynamics and Electroweak Symmetry
  Breaking},''\href{http://dx.doi.org/10.1016/S0370-1573(03)00140-6}{\emph{Phys.
  Rept.} {\bf 381} (2003) 235--402},
  [\href{https://arxiv.org/abs/hep-ph/0203079}{{\tt hep-ph/0203079}}].

\bibitem{Sannino:2016sfx}
F.~Sannino, A.~Strumia, A.~Tesi and E.~Vigiani, ``{Fundamental partial
  compositeness},''\href{http://dx.doi.org/10.1007/JHEP11(2016)029}{\emph{JHEP}
  {\bf 11} (2016) 029}, [\href{https://arxiv.org/abs/1607.01659}{{\tt
  1607.01659}}].
  
  \bibitem{Cacciapaglia:2017cdi}
G.~Cacciapaglia, H.~Gertov, F.~Sannino and A.~E. Thomsen, ``{Minimal
  Fundamental Partial
  Compositeness},''\href{http://dx.doi.org/10.1103/PhysRevD.98.015006}{\emph{Phys.
  Rev.} {\bf D98} (2018) 015006}, [\href{https://arxiv.org/abs/1704.07845}{{\tt
  1704.07845}}].

\bibitem{Alanne:2016rpe}
T.~Alanne, M.~T.~Frandsen and D.~Buarque Franzosi,
``Testing a dynamical origin of Standard Model fermion masses,''
Phys. Rev. D \textbf{94} (2016), 071703, [\href{https://arxiv.org/abs/1607.01440}{{\tt 1607.01440}}].

\bibitem{Belyaev:2016ftv}
A.~Belyaev, G.~Cacciapaglia, H.~Cai, G.~Ferretti, T.~Flacke, A.~Parolini and H.~Serodio,
``Di-boson signatures as Standard Candles for Partial Compositeness,''
JHEP \textbf{01} (2017), 094
[erratum: JHEP \textbf{12} (2017), 088], [\href{https://arxiv.org/abs/1610.06591}{{\tt 1610.06591}}].

\bibitem{Cacciapaglia:2017iws}
G.~Cacciapaglia, G.~Ferretti, T.~Flacke and H.~Serodio,
``Revealing timid pseudo-scalars with taus at the LHC,''
Eur. Phys. J. C \textbf{78} (2018) no.9, 724, [\href{https://arxiv.org/abs/1710.11142}{{\tt 1710.11142}}].

\bibitem{Esteban:2018azc}
I.~Esteban, M.~Gonzalez-Garcia, A.~Hernandez-Cabezudo, M.~Maltoni and T.~Schwetz,
``Global analysis of three-flavour neutrino oscillations: synergies and tensions in the determination of $\theta_{23}$, $\delta_{CP}$, and the mass ordering,''
JHEP \textbf{01} (2019), 106, [\href{https://arxiv.org/abs/1811.05487}{{\tt 1811.05487}}].

\bibitem{Ade:2015xua}
P.~Ade \textit{et al.} [Planck],
``Planck 2015 results. XIII. Cosmological parameters,''
Astron. Astrophys. \textbf{594} (2016), A13, [\href{https://arxiv.org/abs/1502.01589}{{\tt 1502.01589}}].

\bibitem{Tanabashi:2018oca}
{\scshape Particle Data Group} collaboration, M.~Tanabashi et~al., ``{Review of
  Particle
  Physics},''\href{http://dx.doi.org/10.1103/PhysRevD.98.030001}{\emph{Phys.
  Rev.} {\bf D98} (2018) 030001}.

\bibitem{Rosenlyst:2020znn}
M.~Rosenlyst and C.~T.~Hill,
``Natural Top-Bottom Mass Hierarchy in Composite Higgs Models,'' Phys. Rev. D \textbf{101} (2020), [\href{https://arxiv.org/abs/2002.04931}{{\tt
  2002.04931}}].
  
\bibitem{Toma:2013zsa}
T.~Toma and A.~Vicente,
``Lepton Flavor Violation in the Scotogenic Model,''
JHEP \textbf{01} (2014), 160, [\href{https://arxiv.org/abs/1312.2840}{{\tt
  1312.2840}}].

\bibitem{Adam:2013mnn}
J.~Adam \textit{et al.} [MEG],
``New constraint on the existence of the $\mu^+ \to e^+\gamma$ decay,''
Phys. Rev. Lett. \textbf{110} (2013), [\href{https://arxiv.org/abs/1303.0754}{{\tt
  1303.0754}}].

\bibitem{Baldini:2013ke}
A.~Baldini \textit{et al.} [MEG],
``MEG Upgrade Proposal,'' [\href{https://arxiv.org/abs/1301.7225}{{\tt
  1301.7225}}].

\bibitem{Baak:2014ora}
M.~Baak \textit{et al.} [Gfitter Group],
``The global electroweak fit at NNLO and prospects for the LHC and ILC,''
Eur. Phys. J. C \textbf{74} (2014), [\href{https://arxiv.org/abs/1407.3792}{{\tt
  1407.3792}}].

\bibitem{ATLAS:2017ovn}
 ATLAS~collaboration, [ATLAS], ``Combined measurements of Higgs boson production and decay in the $H \rightarrow ZZ^*\rightarrow 4\ell$ and $H\rightarrow\gamma\gamma$ channels using $\sqrt{s}=$ 13 TeV pp collision data collected with the ATLAS experiment,'' ATLAS-CONF-2017-047.

\bibitem{Chen:2013vi}
C.~S.~Chen, C.~Q.~Geng, D.~Huang and L.~H.~Tsai,
``New Scalar Contributions to $h\to Z\gamma$,''
Phys. Rev. D \textbf{87} (2013), [\href{https://arxiv.org/abs/1301.4694}{{\tt
  1301.4694}}].

\bibitem{Kong:2005hn}
K.~Kong and K.~T.~Matchev,
``Precise calculation of the relic density of Kaluza-Klein dark matter in universal extra dimensions,''
JHEP \textbf{01} (2006), 038, [\href{https://arxiv.org/abs/hep-ph/0509119}{{\tt hep-ph/0509119}}].

\bibitem{Kolb:1990vq}
E.~W.~Kolb and M.~S.~Turner,
``The Early Universe,''
Front. Phys. \textbf{69} (1990), 1-547.


\bibitem{Cacciapaglia:2020kbf}
C.~Cai, H.~H.~Zhang, M.~T.~Frandsen, M.~Rosenlyst and G.~Cacciapaglia,
``XENON1T solar axion and the Higgs boson emerging from the dark,''
Phys. Rev. D \textbf{102} (2020) no.7, 075018, [\href{https://arxiv.org/abs/2006.16267}{{\tt
  2006.16267}}].

\bibitem{Aprile:2020tmw}
E.~Aprile \textit{et al.} [XENON],
``Excess electronic recoil events in XENON1T,''
Phys. Rev. D \textbf{102} (2020) no.7, 072004, [\href{https://arxiv.org/abs/2006.09721}{{\tt
  2006.09721}}].

\bibitem{Ahn:2018mvc}
J.~K.~Ahn \textit{et al.} [KOTO],
``Search for the $K_L \!\to\! \pi^0 \nu \overline{\nu}$ and $K_L \!\to\! \pi^0 X^0$ decays at the J-PARC KOTO experiment,''
Phys. Rev. Lett. \textbf{122} (2019) no.2, 021802, [\href{https://arxiv.org/abs/1810.09655}{{\tt
  1810.09655}}].

\end{thebibliography}
\end{document}